\documentclass{aa}  
\usepackage{natbib}
\usepackage{graphicx}
\usepackage{rotating,booktabs}
\usepackage[verbose]{placeins}
\usepackage{pdflscape}
\usepackage{adjustbox}
\usepackage{tikz}
\usepackage[english]{babel}
\usepackage{hyperref}
\hypersetup{colorlinks   = true, citecolor = blue,
            urlcolor=blue,
            linkcolor= blue}

\usepackage{txfonts}
\usepackage{longtable}

\newcommand{\lya}{\ion{Ly$\alpha$}{}}

\newcommand{\citepcantalupo}{\textcolor{blue}{(Cantalupo et al. in prep.)}}

\newcommand{\titu}{\textcolor{blue}{Lazeyras et al. in prep.}}

\defcitealias{Travascio24b}{T24}
\defcitealias{Lepore24}{L24}

\begin{document} 

   \title{X-ray view of a massive node of the Cosmic Web at $z \sim 3$}
   \subtitle{II. Discovery of extended X-ray emission around a hyperluminous QSO}
   
   \titlerunning{Extended hot gas in MQN01}
   
   \author{A. Travascio
          \inst{1}
          \and
          S. Cantalupo
          \inst{1}
          \and
          G. Pezzulli
          \inst{2}
          \and
          P. Tozzi
          \inst{3}
          \and
          L. Di Mascolo
          \inst{2,4}
          \and
          M. Esposito
          \inst{5,6,7,8}
          \and
          T. Lazeyras
          \inst{1}
          \and
          M. Lepore
          \inst{3}
          \and
          S. Borgani
          \inst{5,6,7,8}
          \and
          M. Elvis
          \inst{9}
          \and
          G. Fabbiano
          \inst{9}
          \and
          F. Fiore
          \inst{7}
          \and
          M. Galbiati
          \inst{1}
          \and
          N. Ledos
          \inst{1}
          \and
          R. Middei
          \inst{9,10,11}
          \and
          A. Pensabene
          \inst{1}
          \and
          E. Piconcelli
          \inst{10}
          \and
          G. Quadri
          \inst{1}
          \and
          F. Vito
          \inst{12}
          \and
          W. Wang
          \inst{1}
          \and
          L. Zappacosta
          \inst{10}
          }

\institute{Dipartimento di Fisica "G. Occhialini", Università degli Studi di Milano-Bicocca, Piazza della Scienza 3, I-20126 Milano, Italy \\ e-mail: \texttt{andrea.travascio@unimib.it} \and 
    Kapteyn Astronomical Institute, University of Groningen, Landleven 12, 9747 AD Groningen, The Netherlands \and
    INAF - Osservatorio Astrofisico di Arcetri, Largo E. Fermi 5, 50127 Firenze, Italy \and
    Universit\'e C\^ote d'Azur, Observatoire de la C\^ote d'Azur, CNRS, Laboratoire Lagrange, France \and
    Dipartimento di Fisica - Sezione di Astronomia, Università di Trieste, Via Tiepolo 11, 34131 Trieste, Italy \and
    IFPU, Institute for Fundamental Physics of the Universe, Via Beirut 2, 34151 Trieste, Italy \and
    INAF - Osservatorio Astronomico di Trieste, Via G. B. Tiepolo 11, 34143 Trieste, Italy \and
    INFN, Sezione di Trieste, Via Valerio 2, 34127 Trieste, Italy \and
    Harvard-Smithsonian Center for Astrophysics, 60 Garden St., Cambridge, MA 02138, USA \and
    INAF - Osservatorio Astronomico di Roma, Via Frascati 33, 00040 Monte Porzio Catone, Rome, Italy \and
    Space Science Data Center, Agenzia Spaziale Italiana, Via del Politecnico snc, 135 Roma, Italy
    \and
    INAF - Osservatorio di Astrofisica e Scienza dello Spazio di Bologna, Via Piero Gobetti 93/3, 40129 Bologna, Italy}


\abstract
{While the warm ($T \sim 10^4~\rm K$), ionized gas in the Circumgalactic Medium (CGM) at $z>3$ is now routinely observed around bright quasars in Ly$\alpha$ emission, little is known about the CGM hot phase ($T\sim 10^7~\rm K$), often referred to as the Intracluster Medium (ICM). Here, we report the analysis of 634 ks of \textit{Chandra} X-ray observations in the MQN01 Cosmic Structure, a region containing one of the brightest Ly$\alpha$ nebulae and the largest galaxy overdensity known at $z>3$. We detect $\approx 66$ net counts ($\approx 8 \sigma$) of X-ray emission in the 0.5-2 keV band extending to at least $\sim 30$ kpc from the brightest quasar in MQN01. The morphology and spectrum are consistent with thermal emission from hot plasma in collisional ionization equilibrium. QSO photoionization is found to have a negligible impact, and alternative scenarios such as inverse Compton scattering are disfavored. A joint spatial and spectral MCMC analysis provides consistency with a $\beta$-model with a steep density profile ($\beta \simeq 2.1$, $r_{\mathrm{core}} \simeq 36$ kpc) and an average gas temperature of $kT \simeq (1.8 \pm 0.4)~\rm keV$, corresponding to a virial halo mass of $M_{\mathrm{vir}} \simeq (3 \pm 1) \times 10^{13}~M_{\odot}$. The inferred hot gas mass within the virial radius is
$M_{\mathrm{hot}} \approx 2.6_{-0.6}^{+1.7} \times 10^{12},M_{\odot}$, which is
$\simeq 8.3_{-3.0}^{+9.8}\%$ of $M_{\mathrm{vir}}$, or
$56_{-20}^{+65}\%$ of the theoretical cosmological baryon budget of the halo.
The hot gas also emits an exceptionally high X-ray luminosity, with a measured $L_{2-10} = 2.25_{-1.38} ^{+0.77} \times 10^{45}~\rm erg~s^{-1}$ within the central 30 kpc. This system is a clear outlier in the $L_X$-$T_X$ plane, indicating a thermodynamic state distinct from that of evolved lower-redshift hot halos.  
The cooling time in the inner 15-30 kpc is comparable to the local dynamical time ($1<t_{\mathrm{cool}}/t_{\mathrm{ff}} < 10$) suggesting that the gas could become locally unstable in the absence of heating or feedback. Moreover, the thermal pressure associated with the detected CGM hot phase is large enough to confine the cold and dense clumps, which are required to reproduce the high Ly$\alpha$ emission associated with the inner regions of the MQN01 structure. Although limited to a single system, our results provide unique information on the multi-phase properties of the CGM (or proto-ICM) and a view of the nascent thermal hot gas phase observed in local galaxy clusters.}

\keywords{Galaxies: clusters: intracluster, circumgalactic medium -- X-rays: galaxy: halo -- quasars: supermassive black holes}

   \maketitle

\section{Introduction}

Galaxy clusters host the largest reservoirs of hot gas known as the Intracluster Medium (ICM), which is prominently observed through X-ray emission. According to the hierarchical structure formation paradigm, these massive systems form through the gravitational collapse and merging of smaller structures \citep{Peebles80,NFW96}. Gas infalling into the cluster potential well undergoes shock heating, reaching temperatures of $\sim 10^7-10^8 ~\rm K$ \citep{Rosati02,Kravtsov12}. The dominant X-ray emission mechanism of the ICM is thermal bremsstrahlung, producing luminosities from $L_X \sim 10^{43}$ to more than $10^{45}\rm ~erg~s^{-1}$. 

At higher redshifts ($2<z<3$), massive halos are expected to host a nascent hot-Circumgalactic Medium (CGM) or proto-ICM, influenced by both gravitational processes and non-gravitational effects such as AGN and supernova feedback, although the formation scenario is still unclear \citep{Shimakawa18,Kooistra22}.  
Detecting extended X-ray emission from these structures is particularly challenging due to instrumental sensitivity limits and the need for high angular resolution to separate diffuse emission from point sources. This is crucial at $z>2$, where massive halos often host a central AGN that can dominate the X-ray signal.
Models and simulations predict that the X-ray emission from hot gas at $z>2$ is faint and spatially compact, with an extent on the order of a few tens of arcseconds \citep[i.e., $\sim$tens kpc;][]{Saro09}. Therefore, observations require not only high sensitivity but also high angular resolution, a capability currently achieved only by the \textit{Chandra} X-ray telescope (the mirror FWHM=0.2$\arcsec$ at 1.4 keV) and the ability to perform spatially resolved spectral analysis. Therefore, \textit{Chandra}'s capabilities are unparalleled for studying extended X-ray emission on scales from kpc to a few Mpc, particularly in high-z systems.
Constraining the thermal properties of the CGM/proto-ICM at these early epochs is essential for testing models of the redshift evolution of galaxy halos, such as the transition between cold and hot accretion modes \citep[e.g.,][]{Dekel06}, and for understanding the role of feedback in shaping the thermodynamic state of diffuse gas in forming clusters.

The most distant potential X-ray detection of thermal emission from the ICM associated with a relaxed galaxy cluster has been reported in a system at $z \sim 2.5$ by \citet{WangElbaz16}. A robust detection of thermal ICM emission was observed in the galaxy cluster XLSSC122 at $z \simeq 2$ by \citet{Marrewijk24}. At redshifts $z > 3$, most of the extended X-ray emission has been found around radio-loud galaxies \citep{Yuan03, Smail12}, and is thus associated with non-thermal emission from inverse Compton (IC) scattering of cosmic microwave background (CMB) photons by relativistic electrons in radio jets, as confirmed by the X-ray hard spectrum and the overlap with the radio jets. 
Recently, \cite{Tozzi22b} and \cite{Lepore24} (hereafter \citetalias{Lepore24}) have reported direct evidence of proto-ICM at high redshift, presenting a detailed analysis of the extended thermal X-ray emission within 150 kpc of a radio galaxy in the Spiderweb protocluster at $z = 2.156$. After accounting for non-thermal IC emission from the radio jet, they identified diffuse thermal emission from hot ($kT \sim 2$ keV) gas with a mass of $\sim 1.6 \times 10^{12}~\rm M_{\odot}$, within a halo whose total mass is estimated to be $\sim 0.6-1.4 \times 10^{13}~\rm M_{\odot}$. The combination of X-ray and Sunyaev-Zeldovich \citep[SZ;][]{DiMascolo23} detections revealed an entropy profile indicative of a cool-core system, with a cooling time of less than 100 Myr and a potential mass deposition rate of 250-1000 $\rm M_{\odot}~yr^{-1}$, consistent with infrared-based star formation rates of the central galaxy. These results suggest that AGN feedback and cooling flows may coexist in the early stages of protocluster formation. \\

This work is the second in a series of papers investigating deep (634 ks) \textit{Chandra} X-ray observations of a protocluster at $z=3.25$, centered on the luminous quasi-stellar object (QSO) CTS G18.01 \citep[hereafter ID1; see][]{Travascio24b}. This QSO lies at the center of the Multi Unit Spectroscopic Explorer (MUSE) Quasar Nebula 01 (MQN01), a giant ($> 200 ~\rm kpc$) \lya\ nebula in the sample of \cite{Borisova16}. Recent observations, based on a mosaic of eight VLT/MUSE AO-WFM pointings, have further extended its detection beyond two arcminutes \citepcantalupo. 
To investigate the connection between the properties of this cosmic web node and its galaxy population, multi-wavelength campaigns are actively characterizing the protocluster members \citep{Pensabene24,Galbiati24}. 
In our previous paper, we conducted a census of X-ray AGNs embedded in the protocluster, identifying six AGNs within an area of $\approx 16 \rm ~cMpc^2$ and a velocity range of $\pm 1000~\rm km~s^{-1}$ from the redshift of the central ID1. This corresponds to a signifcant overdensity of $log(L_{2-10~\rm keV/\rm erg~s^{-1}}) > 43$ X-ray AGN relative to the X-ray AGN space density in the field as constrained by \cite{Gilli07}. 

In this second paper, based on the same \textit{Chandra} X-ray dataset, we focus on the extended X-ray emission around the luminous QSO ID1. 
In Section~\ref{sec:methods}, we describe the Chandra data reduction and astrometric correction, respectively. In Section~\ref{sec:detection}, we report a significant detection of spatially resolved X-ray emission at observed energies below 2 keV, extending out to at least 30 kpc from the QSO. Section~\ref{sec:morphology} examines the morphology of this emission in comparison with the \lya\ nebula previously identified around the same QSO by \citet{Borisova16}, while Section~\ref{sec:spectralanalysis} presents an analysis of its spectral properties. In Section~\ref{sec:TB}, we investigate the physical properties of the hot gas under the assumption of a thermal origin. Discussions about the potential influence of QSO photoionization on the thermal emission and alternative emission mechanisms are presented in Sections~\ref{sec:impactQSO} and \ref{sec:alternativemechanisms}. Finally, in Sections~\ref{sec:compSpiderweb} and \ref{sec:ComparisonClusters}, we compare the properties of this system with those of the hot halo in the Spiderweb protocluster, with predictions from numerical simulations, and with observations of local galaxy clusters and groups.

Throughout this paper, all energies are reported in the observed frame unless stated otherwise, with fluxes consistently referring to observed energy bands. In contrast, luminosities are given in rest-frame energy bands. We adopt the same cosmological model as \cite{Travascio24b}\footnote{$H_0 = 67.4 \rm ~km ~s^{-1}~Mpc^{-1}$, $\Omega_{\Lambda} = 0.714$, and $\Omega_m = 0.286$}, in which 1$\arcsec$ corresponds to 7.663 physical kpc. In Section~\ref{sec:TB}, we assume a cosmic baryonic fraction of $f_b = \Omega_b / \Omega_m = 0.15$ to evaluate the reliability of our best-fit model. Unless otherwise noted, all uncertainties shown in the plots correspond to the 1$\sigma$ (68$\%$) confidence level.

\section{Methods}\label{sec:methods}
\subsection{Chandra observations and data reduction} \label{sec:datareduction}

For this study, we used X-ray observations taken with the \textit{Chandra} Advanced CCD Imaging Spectrometer (ACIS-I) during Cycle 23 (2022/2023), with a total exposure time of 634 ks \cite[PI: S. Cantalupo; see][for further details]{Travascio24b}. These observations were conducted in Very Faint (VFAINT) mode, which optimizes the separation of good and bad X-ray events by using the $5 \times 5$ pixel event island for improved event classification. We performed the data reduction using CIAO 4.17 and the \textit{Chandra} Calibration Database (CALDB 4.12.2) installed on Python 3.10. We removed flares identified in the light curves and corrected the astrometry of each ObsID according to the coordinates of the brightest QSO extracted in the GAIA catalog. 

\subsection{Refined astrometric alignment of ObsIDs}\label{sec:align}

In \cite{Travascio24b}, the different ObsIDs were aligned by cross-matching the positions of bright, point-like sources and estimating the least-squares minimization to determine the optimal transformation relative to the deepest ObsID. This is the best method to produce a catalog of X-ray sources in a wide field. However, in this paper, our main purpose is to explore the presence of extended X-ray emission around the AGNs identified in \cite{Travascio24b}. This aim requires a more refined astrometric alignment, made ad hoc for the individual X-ray sources, minimizing positional uncertainties and ensuring that any observed extended emission is not an artifact of residual misalignment. 
However, this approach is feasible only for QSO ID1, which has enough photon counts in each ObsID to reliably determine the centroid position. For the other AGNs, multi-source alignment remains the most suitable option. 
More specifically, for QSO ID1, the 0.5-2 keV image of each ObsID was re-binned to one-fourth of the native pixel size and smoothed with a Gaussian kernel of 3 sub-pixels. The position of the QSO ID1 in each image was determined by computing the centroid's coordinates within a 2$\arcsec$-radius circular region, centered on the approximate location of the QSO. Using the \texttt{wcs\_match} tool, we generated a translation matrix for all ObsIDs, with the deepest ObsID used as the reference. We then applied this transformation to reproject the \texttt{event} and \texttt{aspect} files using the \texttt{wcs\_update} tool.
Figure~\ref{fig:reproj} in Appendix~\ref{app:merged} shows the results of this alignment, displaying the 0.5-2 keV images of the ObsIDs after re-binning and Gaussian smoothing. We then generated a new merged event file by combining the aligned ObsIDs.

\section{Results}\label{sec:results}

\subsection{Detection and validation of extended X-ray emission around the hyperluminous QSO ID1} \label{sec:detection}


\begin{figure*}[t]
   \begin{center}
   \includegraphics[height=0.34\textheight,angle=0]{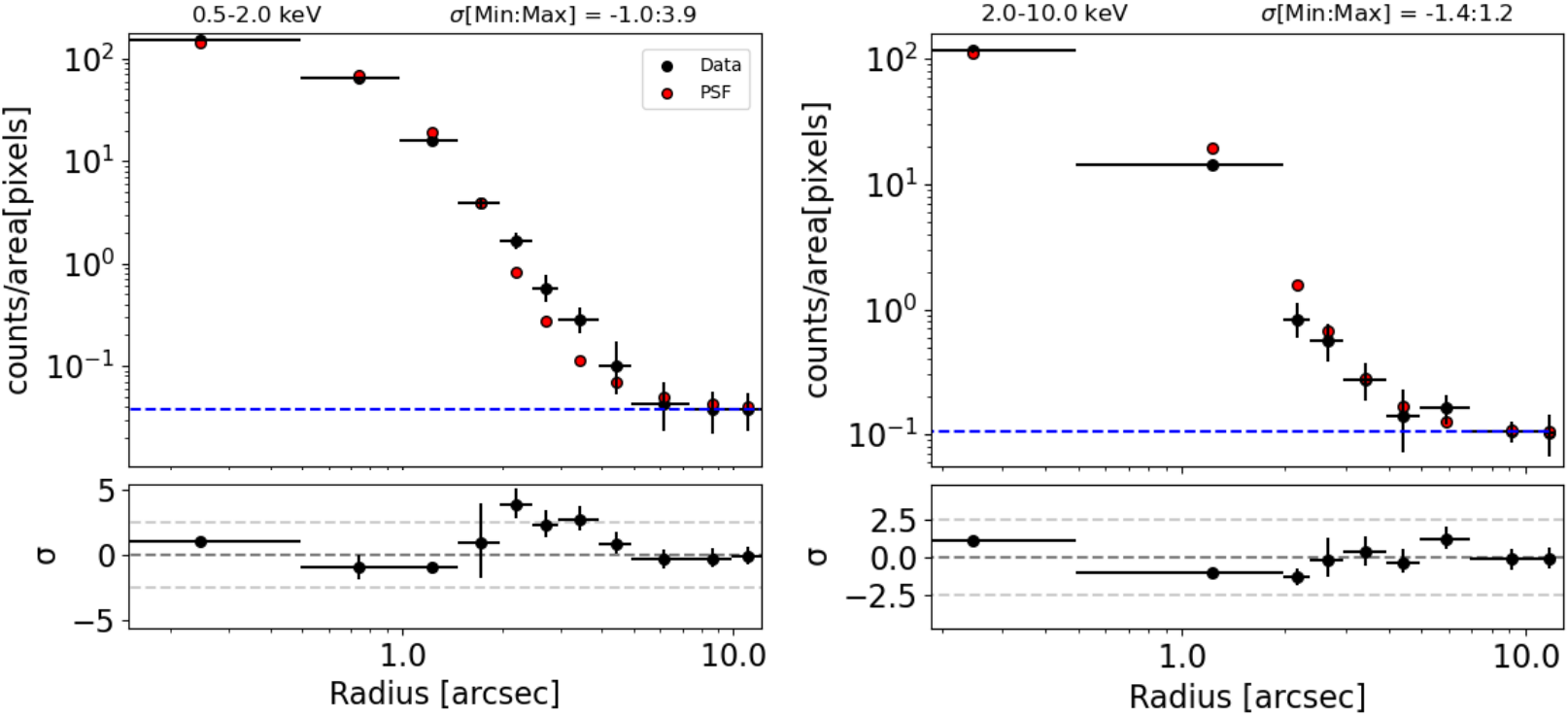}
   \caption{Evidence for residual soft X-ray emission at $<2~\rm keV$ around the brightest QSO in the MQN01 field, ID1, assuming all emission within the central $2 \arcsec$ is due to the AGN. This results in a conservative estimate of the extended component, which is likely non-zero even within this region. Radial profiles of surface counts centered on the QSO ID1, derived from the data (black dots) and the simulated PSF+background (red dots). Profiles are presented for two observed energy bands: 0.5-2 keV (left), and 2-10 keV (right). Radial bins are defined such that each contains sufficient counts to achieve S/N~$>3$. The lower panels display residuals, expressed in units of $\sigma$, as a function of radial bin. The blue dashed line marks the background level, and the horizontal dashed lines in the residual panels correspond to the $-2\sigma$, $0\sigma$, and $+2\sigma$ levels.}
   \label{fig:rp}
   \end{center}
\end{figure*}

\begin{figure*}[t]
   \begin{center}
   \includegraphics[height=0.25\textheight,angle=0]{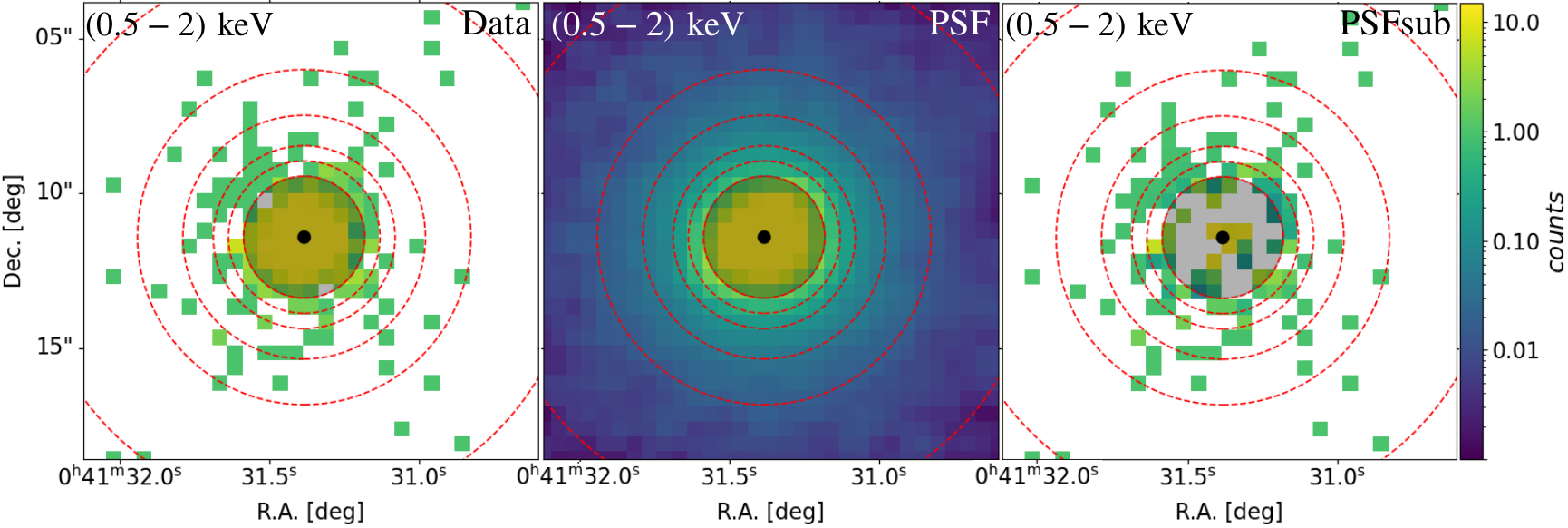}
   \caption{Data (left), simulated PSF (middle), and PSF-subtracted (right) images at the 0.5-2 keV energy band. Red dashed circles indicate the radial bins used for profile extraction, corresponding to the following radii: $\sim$2$\arcsec$,  2.5$\arcsec$,  3$\arcsec$,  4$\arcsec$,  5.5$\arcsec$,  8.9$\arcsec$, 10.3$\arcsec$, and 11.3$\arcsec$. The QSO position is marked with a black dot, and the black shaded circle represents the 2$\arcsec$ radius used to normalize PSF counts to the data. The annular region between 2$\arcsec$ and 4$\arcsec$ may include approximately 6 background counts.}
   \label{fig:rp2}
   \end{center}
\end{figure*}

We investigated the presence of extended X-ray emission around the six X-ray-detected AGN in the MQN01 protocluster \citep{Travascio24b}. For this purpose, we compared the observed emission with the expected point spread function (PSF), considering different energy bands. 
At the position of each AGN and in each selected energy band, we simulated 500 individual PSFs predicted in each ObsID in our dataset. 
To generate these 500 simulated PSFs, we used the \texttt{simulate\_psf} tool\footnote{To model the PSF, we tested blur parameter values ranging from 0 to the default value of 0.25. The blur parameter is meant to account for detector effects and uncertainties in aspect reconstruction that affect the extraction region. We found that variations in the blur parameter did not significantly impact our results, and we adopted a value of 0 for the final PSF modeling.} in CIAO, which accounts for the instrument's response and observational conditions associated with each ObsID. This method requires a spectral model for the energy band of interest. Therefore, we extracted a spectrum within a 2$\arcsec$ radius for each AGN, as this radius contains more than 98$\%$ of the PSF photons, independent of the spectral model.
For each source and energy band, the composite PSF was constructed by weighting each simulated PSF by the exposure time of the corresponding ObsID and summing them.

After generating the PSF images, we compared their radial surface count profiles with those extracted from the observations. We estimated the radial surface count profile of the central AGN and subtracted the background level, determined from the outermost radial bins. For a consistent comparison with the PSF profile, we normalized the simulated PSF by scaling its total counts within a 2$\arcsec$ radius to match the corresponding counts in the observed, background-subtracted count profile centered on the QSO. To mitigate the uncertainties about the PSF centering, we aligned the PSF peak with the centroid of the observed X-ray emission and re-projected the PSF onto the same pixel grid using first-order interpolation. Finally, we quantified any residual excess emission in terms of $\sigma$ by subtracting the expected (PSF+background) counts from the observed counts in each radial bin and dividing by the combined uncertainty, which includes both statistical (Poisson) errors and background estimation uncertainties.

Our analysis reveals a significant extended X-ray emission associated with only one AGN in MQN01, the QSO ID1, in the soft X-ray regime (0.5-2 keV observed, corresponding to $\sim$2-8.5 keV rest-frame at $z=3.2502$). Figure~\ref{fig:rp} displays the radial profiles and residuals for ID1 in the 0.5-2.0 keV (left panel) and 2.0-10.0 keV (right panel) energy bands, extending out to $\sim$12$\arcsec$ ($\approx 90~\rm kpc$), by masking the nearest X-ray neighbors of the QSO, which are ID3 and ID4 in \cite{Travascio24b}. The radial bins were chosen to ensure a minimum S/N of 3 in the data, except for the first bin, which has a fixed radius of 2$\arcsec$. 
Significant extended X-ray emission below 2 keV is observed at $\approx$15-30 kpc (2$\arcsec$-4$\arcsec$) from QSO ID1, with a significance of $\sim$2.5$\sigma$ per radial bin. For comparison, Figure~\ref{fig:nores} in Appendix~\ref{app:nores} presents radial profiles and residuals for AGNs ID2, ID5, and ID6 in the 0.5-2.0 keV and 2.0-10.0 keV bands, where residuals remain below 2$\sigma$, consistent with PSF predictions. AGN ID3 and ID4 were excluded from this analysis due to blending, which prevents a reliable assessment of the presence of extended emission. 

Figure~\ref{fig:rp2} shows the 0.5-2 keV X-ray images of QSO ID1: the observed data (left), the simulated and rescaled PSF (middle), and the PSF-subtracted map (right). Red dashed circles indicate the radial bins used for the extraction of the count profile in Figure \ref{fig:rp}. The PSF was scaled such that its integrated counts within the central 2$\arcsec$ matched those of the observed data. The rightmost panel reveals the spatial distribution of residual counts beyond 2$\arcsec$. A similar image for AGN ID2 is shown in Figure~\ref{fig:imagesID2datapsf} (Appendix~\ref{app:nores}).\\

The detection of extended X-ray emission around the QSO ID1 was confirmed using both the standard merged event file, aligned with multiple reference sources (\citealt{Travascio24b}), and the newly merged event file obtained via the refined alignment procedure described in Section~\ref{sec:align} and Appendix~\ref{app:merged}. Both methods yielded consistent results, though the latter provided a more conservative detection, reducing excess counts within the 2$\arcsec$-4$\arcsec$ annulus by $\lesssim 10 \%$. 
To assess potential PSF estimation errors due to spatial dependencies of the PSF and possible misalignments, we repeated the calculation assuming PSF models estimated from different locations in the field of view. 
We found that only PSF models generated at positions offset by more than $\sim$12$\arcsec$ from the QSO location produced a noticeable impact on the radial profiles. This is far larger than any realistic astrometric misalignment in our data (see, e.g., Figure~\ref{fig:reproj}).

Additionally, we tested an alternative approach in which the PSF was directly subtracted from each ObsID's 0.5-2 keV image before merging, eliminating any potential biases from exposure-weighted PSF stacking. The final PSF-subtracted image remained unchanged, further confirming the robustness of our detection and PSF modeling. 

After subtracting the background and the PSF from the QSO image in the 0.5-2 keV band $\sim 66 \pm 8$ counts remain, indicating a detection with a significance of $\sim 8 \sigma$. This estimate should be considered a lower limit on the extended emission, as it assumes no diffuse component within the central 2$\arcsec$ (i.e., $\sim$15 kpc).

\subsection{Morphology of the extended X-ray emission}\label{sec:morphology}


\begin{figure}[t]
   \begin{center}
   \includegraphics[height=0.28\textheight,angle=0]{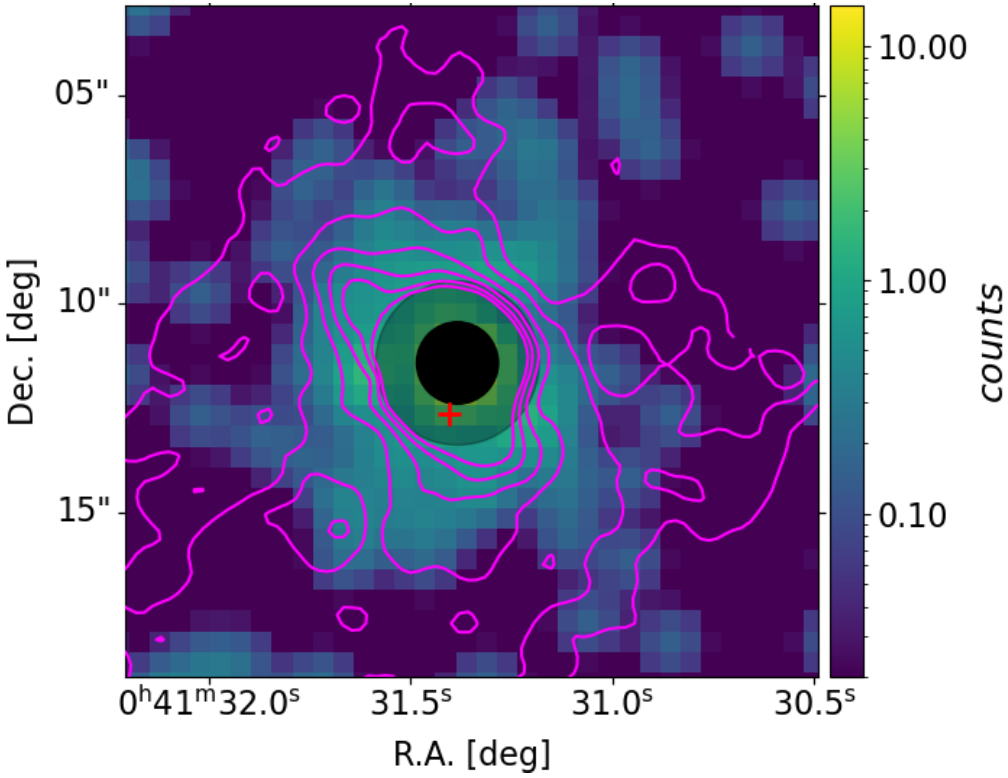}
   \caption{Smoothed soft X-ray 0.5-2.0 keV count map of the extended X-ray emission, obtained after subtracting the QSO's PSF contribution. The filled black dot marks the QSO center with a radius of 1$\arcsec$, while the transparent dot represents the inner 2$\arcsec$ region, where counts are used to rescale the PSF to the image. The magenta contours trace the \lya\ nebula at surface brightness (SB) levels of 2.5, 4, 6, 8, 10, and $12 \times 10^{-18}\rm ergs^{-1}~cm^{-2}~arcsec^{-2}$. The red cross marks the position of the QSO companion detected with ALMA.}
   \label{fig:morphologyhotcold}
   \end{center}
\end{figure}

Around the QSO ID1, there is also evidence of warm ($T \sim 5 \times 10^4~\rm K$) line-emitting gas in the CGM, as indicated by the \lya\ nebula (i.e., MQN01) identified by \cite{Borisova16}, which extends over $200~\rm kpc$. 
If the extended X-ray emission traces hot ($T > 10^6~\rm K$) halo gas, this system provides a unique high-redshift case for multi-phase CGM studies, particularly noteworthy given the absence of a detected radio jet, which rules out jet-driven excitation of the \lya\ emission. 
This provides a remarkable example of the potential coexistence of warm and hot gas phases.
In this section, we focus on comparing the morphology of the extended \lya\ and X-ray emission. Future work will involve a detailed analysis of their physical properties, exploring the implications for gas mass, phase mass fractions, and the underlying processes involved.

Figure~\ref{fig:morphologyhotcold} shows the smoothed map of the 0.5-2 keV counts, obtained after subtracting the QSO's PSF contribution as outlined above. 
The PSF was rescaled to match the sum of the counts within 2$\arcsec$ from the QSO's center, which is marked with a black transparent area, while the black filled dot is centered on the QSO with a 1$\arcsec$ radius. The map was smoothed using the \texttt{aconvolve} tool in CIAO-4.15 with a Gaussian kernel of size $2 \times 2$ pixels, a normalization factor of 1, and a standard deviation of 1 pixel. The magenta contours trace the \lya\ nebula at surface brightness (SB) levels of 2.5, 4, 6, 8, 10 and $12 \times 10^{-18} \rm ~erg~s^{-1}~cm^{-2}~arcsec^{-2}$, providing a direct visualization of the nebular extent as related to the X-ray extended emission morphology. 
 
We observe that the majority of the soft X-ray counts are confined within the \lya\ SB level of $\approx 6 \times 10^{-18}\rm ~erg~s^{-1}~cm^{-2}~arcsec^{-2}$, and seem to be distributed isotropically, with no preferential directions. 
The extended X-ray morphology appears to follow the south tail of the \lya\ contours, while the east \lya\ extension is not connected to any specific X-ray feature. 
The companion of the QSO ObjB, reported in \cite{Pensabene24}, is located approximately $1\arcsec$ from the QSO (red cross in Figure~\ref{fig:morphologyhotcold}), well below the extension of the X-ray emission. Moreover, the observations do not indicate the presence of blended X-ray sources. \\

\begin{figure*}[t]
   \begin{center}
   \includegraphics[height=0.25\textheight,angle=0]{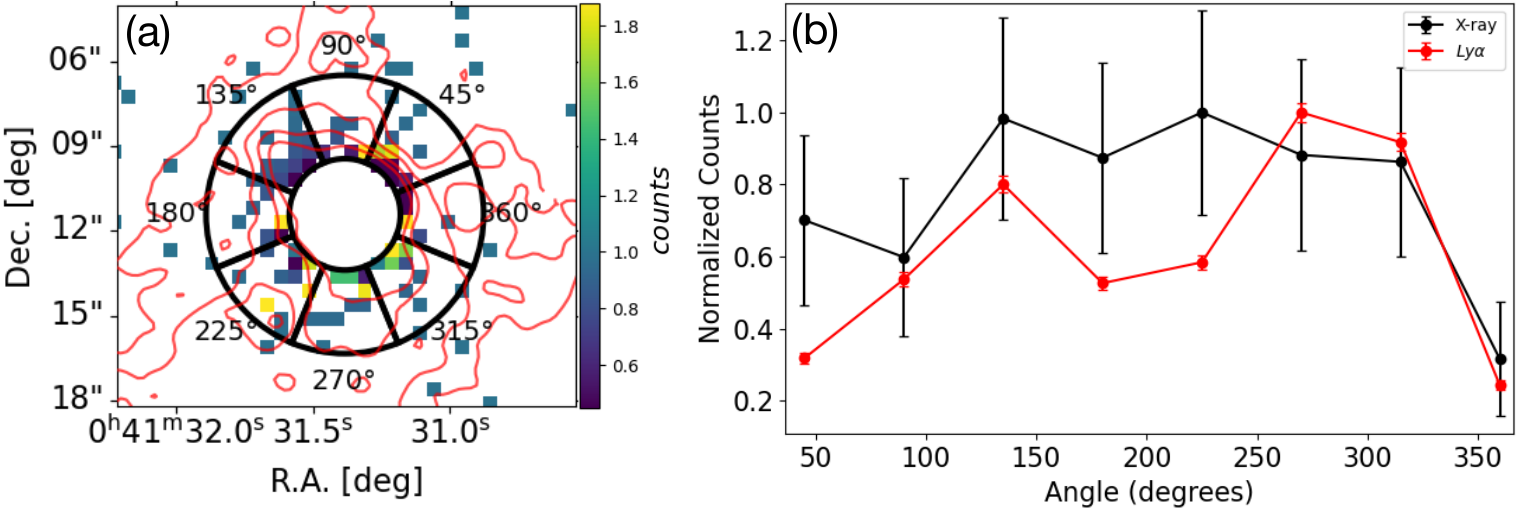}
    \caption{Comparison of the azimuthal flux distribution of the extended \lya\ and X-ray emission. (a) PSF-subtracted map of the 0.5-2 keV \textit{Chandra} X-ray image after subtracting the QSO's PSF. Red contours show the \lya\ SB levels as in Figure~\ref{fig:morphologyhotcold}. Black wedges indicate the sectors used to construct the plot in panel (b). The latter shows the azimuthal distribution of the normalized flux in each 2$\arcsec$-5$\arcsec$ sector, shown as a function of the sector’s central angle. The red curve traces the extended \lya\ emission, while the black curve represents the 0.5-2~keV X-ray emission.}
   \label{fig:azim}
   \end{center}
\end{figure*}


To investigate the spatial correspondence between the extended X-ray emission and the \lya\ nebula, within a radial range of 2$\arcsec$-5$\arcsec$, we divided the region into eight sectors and calculated the mean flux in each azimuthal sector, effectively tracing the angular distribution of both components across the selected annulus. We adopt eight sectors to balance spatial resolution and statistical robustness, allowing us to detect potential azimuthal variations in both emissions. The resulting azimuthal profile, shown in Figure~\ref{fig:azim}, shows a comparison of the trends in the normalized fluxes of the X-ray (black) and \lya\ (red) emissions. 
While the X-ray emission does not show statistically significant azimuthal variations, we note that both X-ray and \lya\ profiles exhibit a mild flux dip near an azimuthal angle of $\sim 360^\circ$, hinting at a possible, but not statistically robust, spatial alignment. This alignment hints at partial morphological coupling between the hot and cold CGM phases, although the X-ray emission remains largely isotropic. This relative isotropy of the hot gas component is consistent with expectations for a virialized or quasi-virialized halo, whereas the more anisotropic morphology of the \lya\ emission likely reflects the clumpy and filamentary structure of the cooler, denser gas phases. 
To quantitatively assess the isotropy of the extended soft X-ray emission, we estimated an anisotropy index ($I_{\mathrm{anis}}$) in $N$ angular sectors. This index quantifies the average fractional deviation of the photon counts in each sector from a perfectly uniform distribution and is defined as:
\begin{equation}
	I_{\mathrm{anis}} = \frac{1}{N} \sum_{i=1}^{N} \left| \frac{C_i - \mu}{\mu} \right|,
\end{equation}
where $C_i$ is the number of counts in the $i$-th sector, $\mu$ is the mean count rate across all sectors, and $N$ is the total number of sectors. 
To assess deviations from uniformity in the extended emission, we applied both a $\chi^2$ test and a Kolmogorov-Smirnov (KS) test to the photon counts distributed in angular sectors. In all cases (including different energy bands and sector choices), we found low isotropy indices ($I_{\mathrm{anis}} < 0.4$) and high $p$-values ($p > 0.2$), with the KS test yielding a maximum $D$-statistic of 0.093 and $p$-value$~= 1$. These results consistently indicate there is no statistically significant deviation from isotropy, suggesting that the emission is consistent with being isotropic within current statistical uncertainties.

\subsection{Spectral Analysis of the Extended Emission}\label{sec:spectralanalysis}

\begin{figure}[t]
   \begin{center}
   \includegraphics[height=0.55\textheight,angle=0]{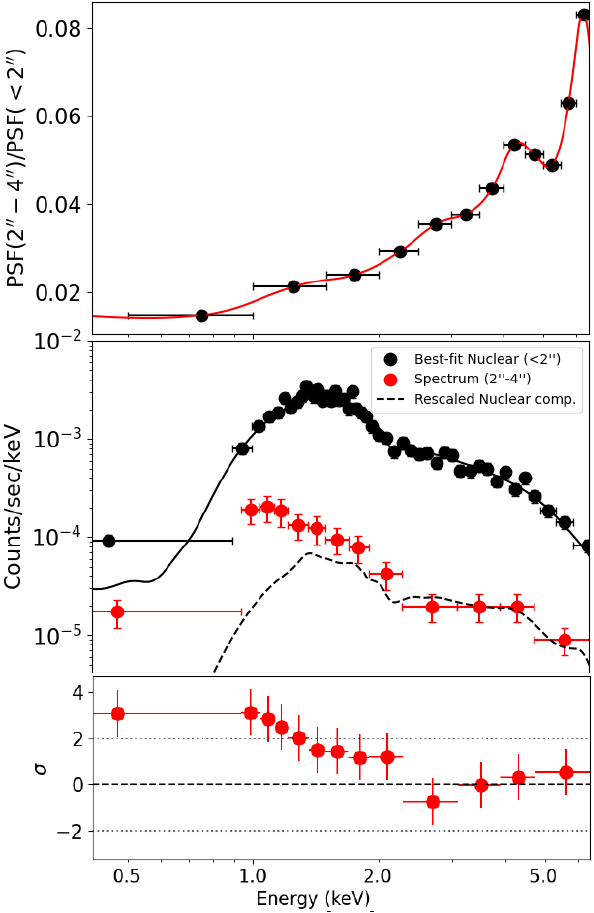}
   \caption{Spectral evidence for excess extended X-ray emission in the soft band, relative to the expected contribution from nuclear (PSF) emission in the 2$\arcsec$-4$\arcsec$ annulus. Top panel: Energy-dependent rescaling factors derived from simulated PSFs, used to estimate the nuclear spillover contribution to the annular spectrum. Middle panel: Spectrum extracted from the central 2$\arcsec$ (black points; binned to at least 50 counts per bin) fitted with an AGN component (solid black line), compared to the spectrum extracted from the 2$\arcsec$-4$\arcsec$ annulus (red points; binned to at least 10 counts per bin), and the predicted nuclear spillover (black dashed line). Bottom panel: Residuals $\sigma$ in the annulus after subtracting the nuclear spillover component.}
   \label{fig:spectot}
   \end{center}
\end{figure}

As a complementary way to investigate and characterize the presence of extended soft X-ray emission, we compare in Figure \ref{fig:spectot} two spectra: one extracted from the inner 2$\arcsec$ region, where the emission is dominated by the QSO, and one from the 2$\arcsec$-4$\arcsec$ annular region (corresponding to $\sim$15-30 kpc), where an excess of soft X-ray emission was detected above $2\sigma$ in Figure~\ref{fig:rp}. The nuclear spectrum is of high quality, with a count rate of $4.15 \times 10^{-3}~\rm cts~s^{-1}$ and a total of 2633 counts, allowing robust spectral fitting. The annular region contains a total of 172 counts in the 0.5-10 keV band and 96 counts in the 0.5-2.0 keV band.

We initially fitted the nuclear spectrum, binning it to a minimum of 50 counts per bin, using the C-statistic (\texttt{C-stat}) and a simple power-law model with Galactic absorption \citep[$N_{\rm H,Gal} = 1.15 \times 10^{20}~\rm cm^{-2}$;][]{Travascio24b}. The analysis was performed with \texttt{Sherpa} \citep{Freeman01,Siemiginowska24} and cross-checked with \texttt{XSPEC} \citep{Arnaud96}. 
The best-fit model without intrinsic absorption yields a C-statistic of 41.19 for 41 degrees of freedom, with a photon index $\Gamma = 2.3 \pm 0.1$ and a normalization of $(2.74 \pm 0.36) \times 10^{-5}$. The corresponding $\chi^2$ is 38.50 for 43 bins (null hypothesis probability $p = 0.582$). 
We also tested the inclusion of intrinsic absorption using the \texttt{xszphabs} model.
The fit improves slightly to a C-statistic of 39.75 for 40 degrees of freedom, with a best-fit intrinsic column density of $N_{\rm H,int} < 6.25 \times 10^{22}~\rm cm^{-2}$. The lower bound of $N{\rm H,int}$ is consistent with zero, and the parameter is fixed at the hard lower limit, indicating that the fit does not require intrinsic absorption. The photon index and normalization remain well determined. 
The small $\Delta$C-stat of 1.44 for 1 additional degree of freedom confirms that the inclusion of intrinsic absorption does not significantly improve the fit. Therefore, we adopt the simpler power-law model with Galactic absorption only. No evidence of a 6.4 keV iron emission line was found, consistent with at most a negligible contribution from X-ray reflection off cold material in the accretion disk or torus. 

Then, we estimated the spillover of nuclear emission into the 2$\arcsec$-4$\arcsec$ annulus. To do it, the best-fit nuclear model was rescaled using energy-dependent correction factors derived from simulated PSFs. These factors, shown in the top panel of Figure~\ref{fig:spectot}, represent the ratio of expected PSF counts in the annular region and the central 2$\arcsec$ aperture. 
The bottom panel of Figure~\ref{fig:spectot} shows the nuclear spectrum (black points), its best-fit model (solid line), the annular spectrum (red points), and the rescaled nuclear spillover component (dashed line). The annular spectrum was binned to a minimum of 10 counts per bin. The comparison shows significant residuals in the annular spectrum after accounting for PSF contribution, particularly below 2 keV, with deviations reaching approximately the 2-4$\sigma$ level in individual bins. This setup (the AGN model in particular) allows us to perform the spectral analysis of the soft emission.

\section{Thermal model of the extended emission} \label{sec:TB}

\begin{figure*}[t]
   \begin{center}
   \includegraphics[height=0.75\textheight,angle=0]{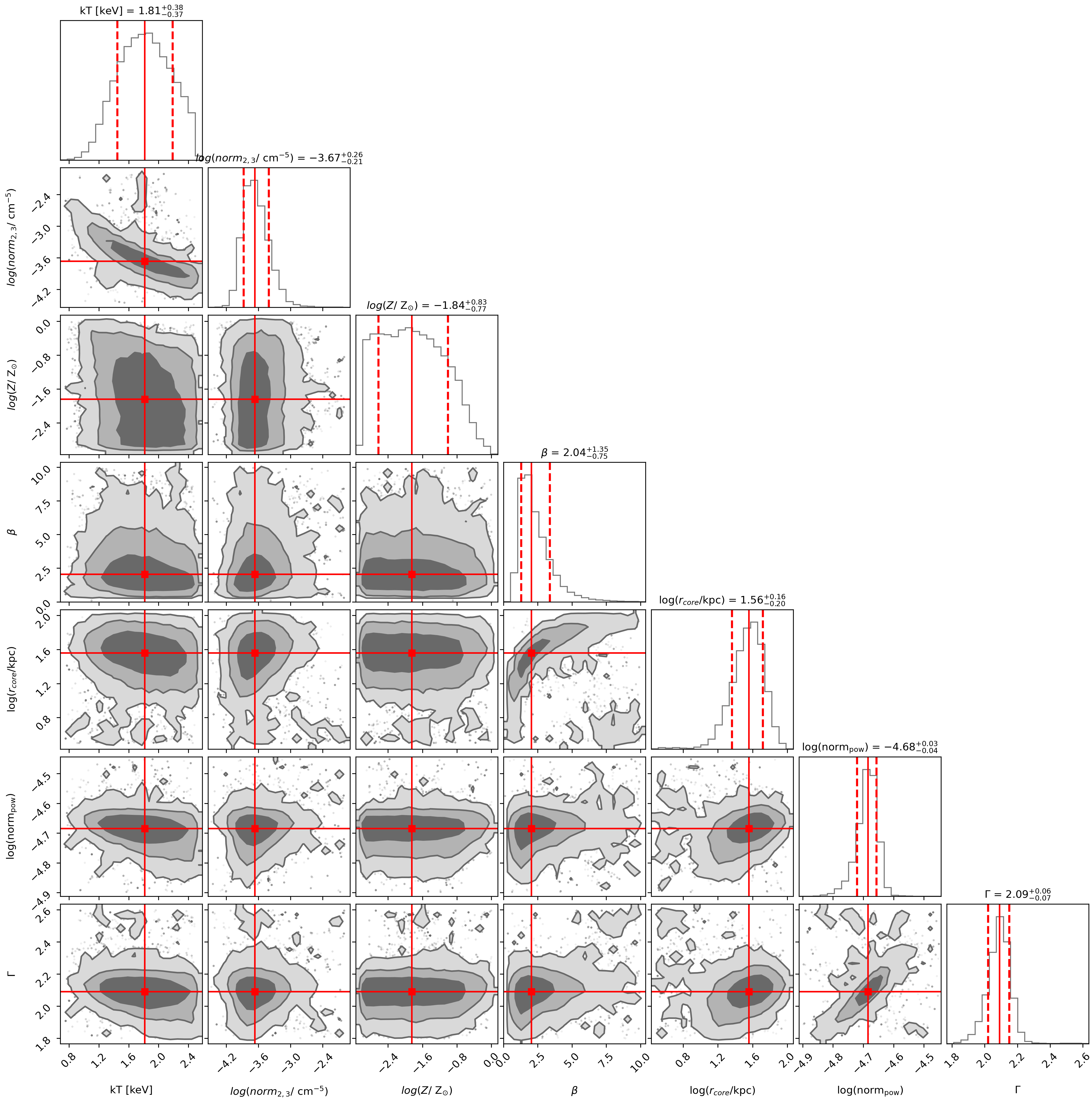}
   \caption{Posterior probability distributions for the simultaneous MCMC modelling of the nuclear and extended emission. Contours represent the 68\%, 95\%, and 99.7\% confidence levels. The solid red vertical lines and the red points mark the best-fit values, which we define as the median value of the distribution of individual parameters, while the red dashed lines mark the 16th and 84th percentiles. Median and percentile values for each parameter are reported above the one-dimensional histograms along the diagonal showing the marginalized distributions for each parameter.}
   \label{fig:MCMCcp}
   \end{center}
\end{figure*}

\subsection{Spectral and spatial analysis of the extended thermal emission} \label{sec:TB2}

In this section, we investigate the physical origin of the spectral excess observed in the X-ray emission (see Section~\ref{sec:spectralanalysis}) and its implications for the hot halo in MQN01. The relatively soft and isotropic spatially extended nature of the emission suggests that it may arise from thermal gas heated to X-ray-emitting temperatures by gravitational shocks (or feedback processes) within the QSO halo. Under this assumption, we model the spectral excess using the \texttt{xsmekal} model in \texttt{Sherpa}, which describes the emission from hot, optically thin plasma in collisional ionization equilibrium (CIE). This model incorporates atomic data for thermal bremsstrahlung, recombination, and line emission from highly ionized species \citep[hereafter referred to as thermal CIE emission; see][]{Mewe85,Mewe86,Liedahl95}. In Section~\ref{sec:alternativemechanisms}, we also show that relaxing the assumption of CIE by including the effect of quasar photo-ionization does not affect our results in any way. 


We performed a joint spectral and spatial analysis of the extended X-ray emission using Markov Chain Monte Carlo (MCMC) methods \citep[e.g.,][]{Ruppin21}, implemented via the Python package \texttt{emcee} \citep{Foreman-Mackey13}. For the comparison between the models and the data, we used four spectra extracted from concentric regions: the central $2\arcsec$, and the annuli spanning $2\arcsec$-$3\arcsec$, $3\arcsec$-$5\arcsec$, and $5\arcsec$-$8\arcsec$. The central spectrum was binned to a minimum of 50 counts per bin, while the others were binned to a minimum of 5 counts per bin. These were modeled using a \texttt{xsmekal} model for the extended emission and a power-law component for the nuclear one (i.e., the PSF). 

As input parameters of the MCMC fitting, we allow the plasma temperature ($kT$), normalization ($norm$), and metallicity ($Z/Z_{\odot}$) to vary freely, adopting solar elemental abundances from the \texttt{Abundanc} table in \cite{Asplund09}. 
We also vary the normalization ($\texttt{norm}_{\rm pow}$) and photon index ($\Gamma$) of the power-law component. We did not include intrinsic absorption for this nuclear component, based on the results presented in Section~\ref{sec:spectralanalysis}, where the column density $N_H$ is consistent with zero within $1.5\sigma$ when fitting the AGN spectrum alone. Nevertheless, even when we include an extended thermal component and allow $N_H$ to be a free parameter, the resulting best‑fit values remain consistent with the current ones within 1$\sigma$, although $N_H$ itself remains close to the lowest detectable limit in our data.

The temperature, metallicity, and normalization ($norm_{2,3}$) of the \texttt{xsmekal} model are those used to fit the $2\arcsec$-$3\arcsec$ annular spectrum (red spectrum in Figure~\ref{fig:spectot}). While we assume that the metallicity and temperature are constant across all the annuli, the parameter $norm_{2,3}$ is proportional to the thermal flux from the $2\arcsec$-$3\arcsec$ annulus. It is physically related to the average squared gas density integrated along the line of sight (see Equation~\ref{eq:norm} below). 
To model the spatial distribution of the thermal gas, we assumed a classical $\beta$-model profile \citep{Cavaliere76}, commonly used to describe the ICM in low-redshift galaxy clusters \citep{Mohr99,Dong10,Conte11,Paggi21}. This model is defined as:
\begin{equation}\label{eq:SB}
	\text{SB}(r) = \text{SB}_{0} \left[ 1 + \left( \frac{r}{r_{\text{core}}} \right)^{2} \right]^{-3\beta + 1/2},
\end{equation}
where $\beta$ and the core radius $r_{\text{core}}$ are free parameters. The thermal normalizations for each spectrum were computed by scaling the initial $norm_{2,3}$ parameter as follows:
\begin{equation}
	norm_{r_1,r_2} = norm_{2,3} \times  \frac{\int _{r_1} ^{r_2} \text{r SB(r)} dr}{\int _{2\arcsec} ^{3\arcsec} \text{r SB(r)} dr}.
\end{equation}
Furthermore, we assumed a constant temperature and metallicity at each radius. 
The $norm_{\rm pow}$ parameter was scaled using energy-dependent simulated PSF profiles (upper panel of Figure~\ref{fig:spectot}).

We adopted an exponential prior for $\beta$ ($P(\beta) \propto e^{-\beta/\beta_*}$, with $\beta_* = 1$), motivated by results from local ICM SB profiles, which generally favor values of $\beta$ below or close to unity, with rare exceptions as large as $\sim$4 \citep[see][]{Mohr99,Dong10,Conte11,Paggi21,Xue00,Wise04,Mirakhor22}. A log-uniform (i.e., scale-invariant) prior was assumed for $r_{\text{core}}$, $Z$, and for the normalization parameters ($norm_{2,3}$ and $norm_{\rm pow}$) to allow for efficient exploration of the parameter space. 

To compare the model predictions with the observed counts, we used the Poisson likelihood (aka C-statistics). To fit each spectrum, we include the background, consisting of a power-law continuum and five Gaussian emission lines, obtained by modeling a spectrum extracted from a large off-source region near the QSO. 
We used 100 walkers and 10000 steps per walker, resulting in a total of $10^6$ samples for the posterior distribution.

Figure~\ref{fig:MCMCcp} shows the marginalized 1D and 2D posterior probability distributions for the model parameters derived from the MCMC. The contours represent the 68.3\%, 95.5\%, and 99.73\% confidence levels.  
In the histograms, the red solid line indicates the median value of each parameter distribution, while the red dashed lines mark the 16th and 84th percentiles. The median values are also shown in the 2D projections with a red square. 
As expected, we found a partial degeneracy between the parameters of the \texttt{xsmekal} model ($kT$ and $norm_{2,3}$), between the parameters of the nuclear power-law ($\Gamma$ and $norm_{\rm pow}$) and between those of the beta model ($\beta$, $r_{\text{core}}$). The latter is the most marked, reflecting the absence of deep data at large radii, which would be needed to better constrain the spatial distribution of the gas. 
The degeneracy between $kT$ and $norm_{2,3}$ of the \texttt{xsmekal} model is mostly due to the lack of data at lower energies (at this redshift, we can see with \textit{Chandra} only the exponential tail of the Bremsstrahlung). Despite these limitations, the marginalized 1D posterior probability distributions are all well-behaved and unimodal for all parameters, allowing us to put clear quantitative constraints on the properties of the hot gas. In particular, we found as median temperature and normalization of the thermal component: $kT = (1.8 \pm 0.4)~\rm keV$ (corresponding to $T = (2.1 \pm 0.4) \times 10^{7}~\rm K$), and $norm_{2,3} = 2.13_{-0.82}^{+1.75} \times 10^{-4}~\rm cm^{-5}$. The nuclear power-law parameters are well constrained, with $norm_{\text{pow}} = (2.1 \pm 0.2) \times~10^{-5}$ and $\Gamma = 2.1 \pm 0.1$. Note that the best fit slope of the nuclear component is slightly shallower than (albeit within $2\sigma$ from) that inferred from the analysis of the nuclear spectrum alone (2.3; Section \ref{sec:spectralanalysis}), reflecting a small but not entirely negligible contribution of the thermal emission also at small radii (see also Table \ref{tab:flux_luminosity_unc} below). The $\beta$-model parameters describing the spatial distribution of the hot gas are $\beta = 2.04_{-0.75}^{+1.35}$ and $r_{\text{core}} = 36_{-13}^{+16}~\rm kpc$. A complete list of the best-fit parameters is provided at the top of Table~\ref{tab:parMCMC_singleZ}. 
The only parameter that is loosely constrained by our fit is the metallicity. The posterior probability distribution for this parameter is very broad, reflecting an insufficient S/N for a precise measurement. The posterior probability nonetheless shows a clear preference for relatively low metallicities ($Z \lesssim 0.1 Z_\odot$), mainly as a consequence of the non-detection of the Fe K$\alpha$ complex from He-like and H-like ions at rest-frame 6.7-6.9 keV (observed $\approx$1.6 keV). We stress, however, that metallicity is included in our MCMC fit only for marginalization purposes, and the formal best-fit value reported in Table~\ref{tab:parMCMC_singleZ} should be taken with caution.

\begin{figure*}[t]
   \begin{center}
   \includegraphics[height=0.445\textheight,angle=0]{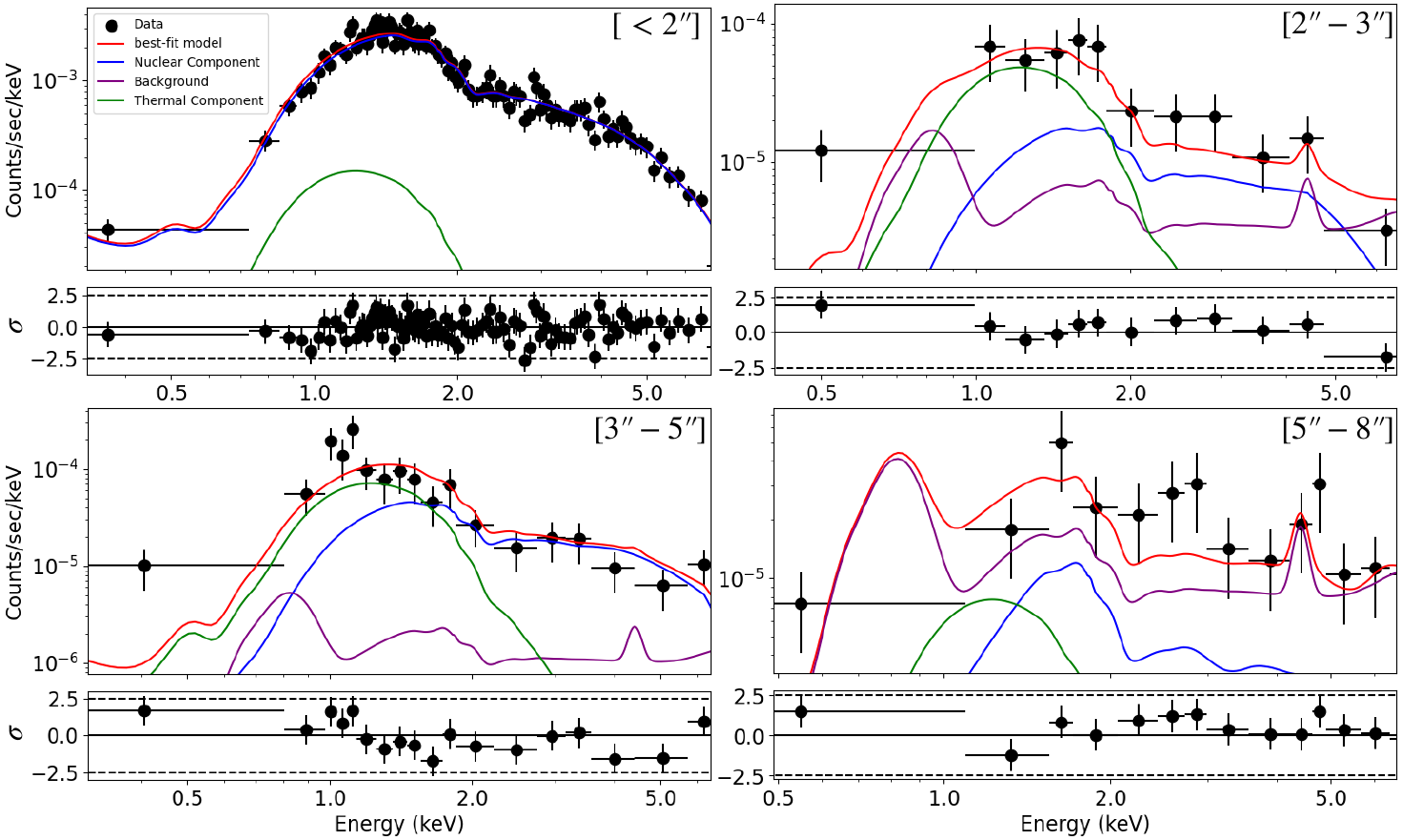}
   \caption{Observed spectra extracted from four regions: the central 2$\arcsec$ aperture (top left), and the 2$\arcsec$-3$\arcsec$, 3$\arcsec$-5$\arcsec$, and 5$\arcsec$-8$\arcsec$ annuli (top right, bottom left, and bottom right, respectively). Overplotted are the best-fit models (red lines) based on the median posterior values from Table~\ref{tab:parMCMC_singleZ}. The individual spectral components are shown in blue (nuclear power-law), green (thermal emission), and purple (background). Residuals in the lower sub-panels are given in units of $\sigma$.}
   \label{fig:spectraMCMC}
   \end{center}
\end{figure*}

Figure~\ref{fig:spectraMCMC} presents the best-fit models (red lines) for the four spectra extracted from the regions defined previously: the central 2$\arcsec$ aperture (top left), and the three concentric annuli 2$\arcsec$-3$\arcsec$, 3$\arcsec$-5$\arcsec$, and 5$\arcsec$-8$\arcsec$, shown in the top-right, bottom-left, and bottom-right panels, respectively. The individual spectral components are plotted separately: the nuclear power-law (blue), background (purple), and thermal (green) components. 
The residuals in the bottom sub-panels of each spectrum are expressed in terms of $\sigma$. The best-fit models reproduce the observed spectra within $\sim 2 \sigma$ across all energy bins. However, in the 2$\arcsec$-3$\arcsec$ and 3$\arcsec$-5$\arcsec$ spectra, the models tend to underpredict the emission in the lowest-energy bin (below 0.7 keV). However, emission below 1 keV may be subject to systematic uncertainties related to calibration accuracy, as discussed in Section~\ref{sec:systematics}.  

\begin{table}[h]
\centering
\caption{Median, 16th, and 84th percentiles of the posterior distributions from the MCMC analysis, along with the derived physical properties of the hot gas halo.}
\vspace{0cm}
\small
\renewcommand{\arraystretch}{1.3}
\setlength{\tabcolsep}{6.pt}
\begin{tabular}{lcc}
\toprule
\textbf{Parameter} & \textbf{Units} & \textbf{Value} \\
\midrule
$kT$ & [keV] & $1.81_{-0.37}^{+0.38}$ \\
$\rm norm$ & [$10^{-4}~\text{cm}^{-5}$] & $2.13_{-0.82}^{+1.75}$  \\
$\rm Z ^{\dagger}$ & $[Z_{\odot}]$ & $0.014_{-0.012}^{+0.083}$  \\
$\beta$ & & $2.04_{-0.75}^{+1.35}$  \\
$r_{\text{core}}$ & [kpc] & $36_{-13}^{+16}$ \\ 
$norm_{\rm pow}$ & [$10^{-5}$] & $2.07_{-0.18}^{+0.15}$ \\
$\Gamma$ & & $2.09_{-0.07}^{+0.06}$  \\ 
\midrule   
$T$ & [$10^7$ K] & $2.1_{-0.4}^{+0.4}$ \\
$M_{\rm vir}$ & [$10^{13}~M_\odot$] & $3 \pm 1$ \\
$R_{\rm vir}$ & [kpc] & $190_{-20}^{+19}$ \\
$n_{\text{e},0}$ & [$\text{cm}^{-3}$] & $0.86_{-0.19}^{+0.48}$ \\
$M_{\rm hot~gas}(R_{\rm vir})$ & [$10^{12}~M_\odot$] & $2.6_{-0.6}^{+1.7}$ \\
$M_{\rm hot}/M_{\rm vir}$ & & $0.083_{-0.030}^{+0.098}$ \\
$f_{\rm hot}$ & & $0.56 _{-0.20}^{+0.65}$ \\
\bottomrule
\end{tabular}
\label{tab:parMCMC_singleZ}
\\[0.2cm]
$^\dagger$ parameter used as marginalization (not constrained)
\end{table}

\subsection{Physical properties of the hot halo gas}\label{sec:prophalo}

We used the median posterior values to estimate several physical properties of the hot gas. Assuming the system is virialized, we derived a virial mass (equivalent to $M_{200}$) of $M_{\text{vir}} \approx 3 \times 10^{13}~\rm M_{\odot}$ from the gas temperature and the redshift of the system, following Equation (A10) from \citet{Dekel06}. This estimate is consistent with scaling relations reported in the literature \citep[e.g.,][]{Sato00, Wang08, Ehlert09}. The corresponding virial radius (i.e., $R_{200}$), calculated assuming the same overdensity criterion, is $R_{\text{vir}} \simeq 190~\rm kpc$, based on Equation (2) from \citet{Ehlert09}. This yields a ratio $r_{\text{core}} / R_{\text{vir}} \approx 0.2$, which falls within the typical range of $10^{-2}$ to $0.6$ observed for the ICM in local galaxy clusters and groups \citep[e.g.,][]{Vikhlinin05}.

The electron and hydrogen density distribution in the plasma is related to the best-fit $norm_{2,3}$ parameter through the expression\footnote{See \url{https://heasarc.gsfc.nasa.gov/docs/xanadu/xspec/manual/node198.html}}:
\begin{equation} \label{eq:norm}
	norm_{r_{\rm in},r_{\rm out}}~[\text{cm}^{-5}]  = \frac{10^{-14} \int _{r_{\rm in}} ^{r_{\rm out}} n_{\text{e}}(r) \, n_{\text{H}}(r) \, dV}{4 \pi [D_A(1+z)]^2},  
\end{equation}
where $D_A$ is the angular diameter distance to the source, and the volume integral includes only the region along the line of sight that contributes to the $r_{\rm in} -r_{\rm out}$ annulus in projection. 
Accounting for the contribution of electrons from helium (0.17 per hydrogen ion, assuming full ionization and a helium mass fraction $Y=0.25$), we adopt $n_{\text{e}}(r) \simeq 1.17 \, n_{\text{H}}(r)$, making the $norm$ parameter proportional to $\int n_{\text{e}}^2(r) \, dV$. This expression depends on the electron density profile, which is inferred by deprojecting the observed SB profile under the assumption that the X-ray emissivity scales with $n_{\text{e}}^2(r)$.  For the adopted $\beta$-model (equation \ref{eq:SB}), the deprojected electron density distribution is
\begin{equation}\label{eq:ne}
	n_{\text{e}}(r) = n_{\text{e},0} \left[ 1+ \left(\frac{r}{r_{\text{core}}} \right)^2 \right]^{-\frac{3\beta}{2}},
\end{equation}
where $n_{\text{e},0}$ is the central electron density.
To relate this 3D density profile to the observed 2D spectral normalization in a given annulus, we perform a proper deprojection of the spherical volume intersected by the line of sight through the annulus. Specifically, we compute the volume integral in Equation~(\ref{eq:norm}) using cylindrical coordinates ($R$,$\phi$,$z$), where the volume element is $dV = R dR\, d \phi \, dz$ and $z=\sqrt{r^2 -R^2}$. The azimuthal integration yields a factor $2\pi$, using the analytical form of the line-of-sight projection for spherically symmetric functions, commonly known from the Abel transform framework, we obtained:
\begin{equation} \label{eq:norm_proj_correct1}
    \int n_{\text{e}}^2(r) dV = \frac{4 \pi r_{\text{core}}^3 n_{\text{e},0}^2 \, \tilde{\chi} (3 \beta)}{\beta_0} \Biggl[ \Biggl( 1 + \frac{r_{in}^2}{r_{\text{core}}^2} \Biggr)^{-\frac{\beta_0}{2}} - \Biggl( 1 + \frac{r_{out}^2}{r_{\text{core}}^2} \Biggr)^{-\frac{\beta_0}{2}} \Biggr] 
\end{equation}
where $\beta_0 = 6 \beta -3$, while the function $\tilde{\chi}$ is defined as: 
\begin{equation} \label{eq:tchia}
    \tilde{\chi}(a) = \int_0^{+\infty} (1+x^2)^{-a} dx = 
\frac{\sqrt{\pi}}{2}\frac{\Gamma(a-\frac{1}{2})}{\Gamma(a)}.
\end{equation}
A proof of the last equality in equation \eqref{eq:tchia} is provided in Appendix \ref{app:eqchi}. Note that $\tilde{\chi}$ is related to the function $\chi$ defined in \cite{{Pezzulli17}} by the relation $\tilde{\chi}(a) = \chi(2a)$.

This formulation (\ref{eq:norm_proj_correct1}) accurately captures the projection of the spherical density distribution into the annular region observed in the sky. By inserting this expression into Equation~(\ref{eq:norm}), we derived a direct relationship between the central density and the observed spectral normalization in the $n_{\text{e},0}$ corresponding $r_{\rm in}$-$r_{\rm out}$ annulus:
\begin{equation} \label{eq:norm_proj_correct2}
n_{\text{e},0} = \sqrt{ \frac{1.17 \times 10^{14}\, [D_A(1+z)]^2 norm_{r_{\rm in},r_{\rm out}} \, \beta_0}{r_{\text{core}}^3 \, \tilde{\chi}(3 \beta) \, [ ( 1 + r_{\rm in}^2/r_{\text{core}}^2)^{-\frac{\beta_0}{2}} - ( 1 + r_{\rm out}^2/r_{\text{core}}^2)^{-\frac{\beta_0}{2}} ]} } 
\end{equation}
where $D_A$ and $r_{\text{core}}$ are expressed in cm, $norm_{r_{\rm in},r_{\rm out}}$ in $\rm cm^{-5}$ and $n_{\text{e},0}$ in $\rm cm^{-3}$.
Applying this method, we estimated a central electron density of $n_{\text{e},0} = 0.9_{-0.2}^{+0.4}~\rm cm^{-3}$. We then computed the total mass of hot gas within the virial radius using:
\begin{equation}
	M_{\text{hot gas}}  = \frac{4 \pi \, n_{\text{H},0} \, m_p }{X} \int _0 ^{R_{\text{vir}}}  r^2 \,  \Biggl[ 1 + \Biggl( \frac{r}{r_{\text{core}}} \Biggr)^2 \Biggr] ^{-3 \beta /2} dr
\end{equation}
where $n_{\text{H},0} = n_{\text{e},0}/1.17$, $m_p = 8.4 \times 10^{-58}~M_{\odot}$, and $X=0.75$ the hydrogen mass fraction. 
This yields a hot gas mass of $M_{\text{hot gas}} (<R_{\text{vir}}) = 2.6_{-0.6}^{+1.7} \times 10^{12}~\rm M_{\odot}$ corresponding to a hot baryon fraction of $M_{\rm hot}/M_{\rm vir} \approx 0.083^{+0.098}_{-0.030}$. Assuming a cosmic baryon fraction $f_b = \Omega_b / \Omega_m =0.15$ \citep[e.g.][]{Mohr99,Gonzalez07,Ettori09,Hinshaw13,Ge18} this implies that the hot gas accounts for $f_{\rm hot~gas} = (M_{\rm hot}/M_{\rm vir})/0.15 = 0.56_{-0.20}^{+0.65}$ of the baryons primordially associated with the halo. This fraction highlights that a substantial part of the halo's theoretical baryon budget is found in the hot CGM phase. Here, by baryon budget, we refer to the total baryonic mass expected from the cosmic baryon fraction to be associated with the halo's dark matter, independent of whether these baryons are currently located inside or outside the virial radius. This fraction is consistent with observations of low-mass groups and intermediate-mass clusters in the local Universe, where the hot gas typically accounts for about 30-85$\%$ of the halo's baryonic content \citep[e.g.,][]{Eckert13b,Morandi15,Pratt23}. 
The fact that a similarly large mass of hot CGM is found already at $z \sim 3$, co-existent with a luminous Giant Ly$\alpha$ nebula, is in line with the predictions of \cite{Pezzulli19}, who explored the role of hot virialized gas in boosting the Ly$\alpha$ emissivity in MUSE QSO nebulae (including MQN01) through the compression of colder (Ly$\alpha$-emitting) gas. This is further discussed in Section~\ref{sec:pressureconfinement}.

\begin{table}[h] 
\centering
\caption{Soft (0.5-2 keV) and hard (2-10 keV) absorbed fluxes (observed energy), and unabsorbed luminosities (rest-frame energy) of the thermal and power-law components in the central and intermediate annuli, including 1$\sigma$ uncertainties. These values are derived assuming the best-fit parameters reported in Table~\ref{tab:parMCMC_singleZ}.}
\vspace{0.2cm}
\small
\renewcommand{\arraystretch}{1.5}
\setlength{\tabcolsep}{1pt}
\begin{tabular}{lccc}
\hline
Regions & Components & $F_{0.5-2~\mathrm{keV}}$ / $F_{2-10~\mathrm{keV}}$ & $L_{0.5-2~\mathrm{keV}}$ / $L_{2-10~\mathrm{keV}}$ \\
        &            & [$\rm erg/cm^{2}/s$] & [$\rm erg/s$] \\
        &            & ($10^{-15}$) & ($10^{44}$)  \\
\hline
0-2$\arcsec$ & thermal    & $5.38^{+1.64}_{-1.14}$ / $0.12^{+0.07}_{-0.05}$ & $12.22^{+4.68}_{-3.32}$ / $6.15^{+2.03}_{-1.29}$ \\
             & power law  & $44.11^{+1.74}_{-1.69}$ / $46.49^{+1.76}_{-1.21}$ & $51.63^{+4.61}_{-4.36}$ / $52.56^{+2.02}_{-1.97}$ \\
\hline
2$\arcsec$-3$\arcsec$ & thermal    & $2.59^{+0.53}_{-0.46}$ / $0.06^{+0.03}_{-0.02}$ & $5.79^{+2.11}_{-1.44}$ / $3.00^{+0.61}_{-0.55}$ \\
                      & power law  & $0.79^{+0.03}_{-0.03}$ / $0.84^{+0.03}_{-0.02}$ & $0.93^{+0.08}_{-0.08}$ / $0.95^{+0.04}_{-0.04}$ \\
\hline
3$\arcsec$-5$\arcsec$ & thermal    & $1.82^{+0.40}_{-0.39}$ / $0.04^{+0.02}_{-0.01}$ & $4.03^{+1.64}_{-1.07}$ / $2.13^{+0.43}_{-0.45}$ \\
                      & power law  & $0.31^{+0.01}_{-0.01}$ / $0.33^{+0.01}_{-0.01}$ & $0.36^{+0.03}_{-0.03}$ / $0.37^{+0.01}_{-0.01}$ \\
\hline
\end{tabular}
\label{tab:flux_luminosity_unc}
\end{table}

Table~\ref{tab:flux_luminosity_unc} summarizes the soft (0.5-2 keV) and hard (2-10 keV) absorbed fluxes and unabsorbed luminosities of the thermal and nuclear components for each spectrum, assuming best fit (median) parameters, excluding the outermost annulus, where the emission lies below the background level. 
The results indicate that, within the central 2$\arcsec$ region, the observed thermal soft X-ray flux may account for approximately $12 \pm 4 \, \%$ of the total emission in the central $2 \arcsec$. However, this estimate depends sensitively on the $\beta$-model parameters.

\subsection{Systematic uncertainties on the X-ray luminosities}\label{sec:systematics}

An important aspect to consider is the presence of systematic uncertainties affecting the measurement of the soft X-ray luminosities (Table~\ref{tab:flux_luminosity_unc}). The detection of extended thermal emission in the $0.5$-$2$~keV rest-frame band may be susceptible to the shape of the spectral continuum at low energies, where the effective area of \textit{Chandra} drops significantly. Since the thermal component is mainly constrained by photons below $\sim2$~keV, any small number of counts or calibration inaccuracy in this range, although unlikely to fully mimic the observed signal, can steepen the spectral fit and bias the extrapolated luminosity estimate.
However, it is worth noting that our main MCMC model already underpredicts the low-energy counts (see residual for the lowest energy bin in Figure~\ref{fig:spectraMCMC}), suggesting that any such systematic would only bring out the model in better agreement with the data. More generally, the mere detection of soft X-ray emission at such high redshift, where the observed energy range samples only the high-energy tail of the thermal bremsstrahlung spectrum, already implies that the emission must be intrinsically powerful.


Nonetheless, to mitigate any unmodelled systematics related to the lowest energy bin, we repeated the MCMC analysis using only the spectral data above $1$~keV (observed frame), where the \textit{Chandra} effective area is more stable and remains above $60~\mathrm{cm}^2$. In this alternative fit, we applied a Gaussian prior on the temperature, centered at $kT = 2.5$~keV (i.e., 64\% higher than the previous best-fit value) with a standard deviation of $0.5$~keV. Imposing a prior with a higher temperature than our fiducial model is a conservative choice, as a higher temperature implies a shallower spectral slope, thereby reducing the extrapolated soft-band luminosity. 
Under these assumptions, the posterior distribution for $kT$ peaks around $\approx 1.9$~keV, with the corresponding total luminosity, within 30 kpc, reaching a minimum of $\approx 10^{45}~\mathrm{erg\,s^{-1}}$ within $1\sigma$. This test shows that, even when conservatively accounting for potential calibration systematics and introducing an explicit bias in favour of high temperatures, relatively low temperatures are still preferred and the inferred intrinsic soft-band luminosity remains significantly high.

\section{Discussion}

\subsection{Impact of the QSO radiation on the thermal emission}\label{sec:impactQSO}

Throughout our X-ray spectral analysis, we have implicitly assumed that the hot gas is in CIE, meaning that its ionization state and emitted spectrum are determined just by collisions with thermal electrons. However, this assumption may break down due to the presence of intense photoionizing radiation from the central QSO. In our case, the QSO hosted in the center of MQN01 protocluster is hyper-luminous, emitting copiously in both the UV and X-ray bands. This extreme radiation field has the potential to alter the ionization balance and emissivity of the surrounding gas, particularly within the central 30 kpc. It is therefore important to test whether AGN photoionization can significantly affect the thermal X-ray emission, and hence bias our interpretation of the physical properties derived under the CIE assumption.

We used \texttt{Cloudy} \citep{Ferland98} to model the emission from hot CGM gas, both with and without the influence of AGN photoionization. Specifically, we simulated a coronal gas as in our fiducial model. For this test, we focused on a shell with radii 15-23 kpc, which corresponds in projection to the $2\arcsec$-$3\arcsec$ annulus where the signal of the extended emission is the strongest (see Figure~\ref{fig:spectraMCMC}), and we assume a temperature and metallicity of $kT = 1.8~\rm keV$ and $Z \simeq 0.014 \, Z_{\odot}$, and a gas density $\langle n_{\text{H}} \rangle \simeq 0.4~\rm cm^{-3}$, which is the average density value in such shell for our fiducial beta model (see Section~\ref{sec:compSpiderweb}). The gas was placed at this distance in a shell geometry within \texttt{Cloudy}.

\begin{figure}[t]
   \begin{center}
   \includegraphics[height=0.61\textheight,angle=0]{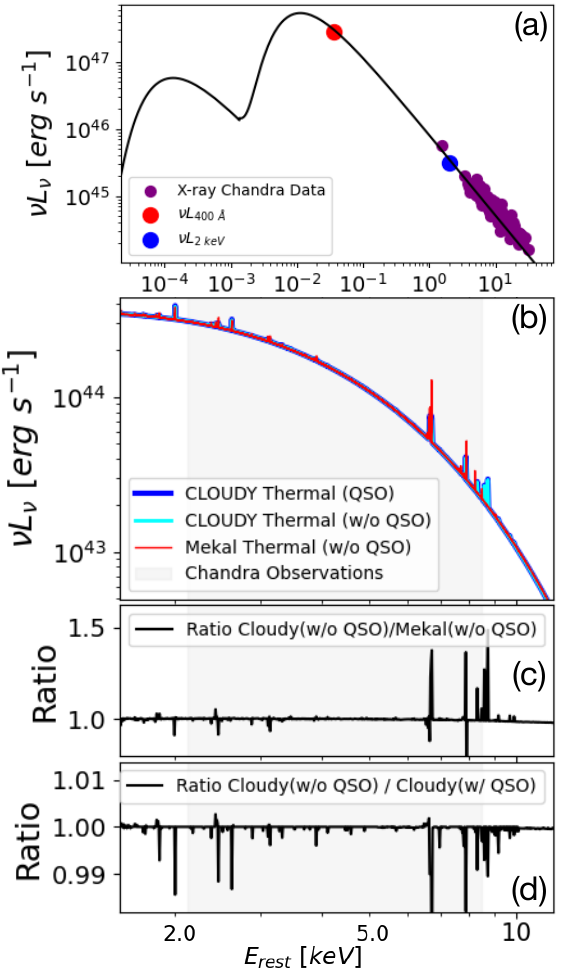}
\caption{Impact of QSO photoionization on the thermal emission in the observed 0.5-2 keV energy band (highlighted by the shaded gray region). Panel (a) shows the SED adopted to represent the QSO radiation field, constrained by two photometric measurements (UV and X-ray; red and blue dots) for this source and by the power-law slope $\Gamma \approx 2.1$ used to match the unfolded \textit{Chandra} X-ray data (purple dots). Panel (b) compares the thermal emission spectrum predicted by the standard \texttt{xsmekal} model (red line) with that obtained from \texttt{Cloudy} simulations including QSO photoionization (see text for details). Panel (c) shows the ratio between the \texttt{Cloudy} model with QSO photoionization and the \texttt{xsmekal} model. Panel (d) displays the ratio between \texttt{Cloudy} thermal models computed with and without the QSO radiation field, illustrating the net impact of photoionization.}
   \label{fig:cloudymekal}
   \end{center}
\end{figure}

The energy injected by the central AGN was described using a spectral energy distribution (SED; panel (a) of Figure~\ref{fig:cloudymekal}) tabulated from observations ("\texttt{AGN T=3e5 K a(ox)=-1 a(uv)=-2 a(x)=-2.2}"), constrained to match UV \citep[$\nu L_{\nu}(400\,\text{\AA}) = 5 \times 10^{46}~\rm erg\,s^{-1}$;][]{Borisova16} and X-ray \citep[$\nu L_{\nu}(2\,\text{keV}) = 3.3 \times 10^{45}~\rm erg\,s^{-1}$;][]{Travascio24b} photometric measurements, adopting a X-ray power law photon-index of $\Gamma \approx 2.1$. This setup allows us to test the potential impact of AGN photoionization on the thermal emission from the surrounding gas. We also ran an additional \texttt{Cloudy} simulation where the QSO SED was suppressed by 6 orders of magnitude, to simulate a condition where photoionization is negligible and CIE applies. This was done to allow for a direct comparison with the \texttt{xsmekal} model (which also assumes CIE) and test the dependence of the results on different codes at fixed physical assumptions.

We then reproduced the corresponding thermal emission using the \texttt{xsmekal} function in \texttt{sherpa}, adopting the same parameters as before: $kT = 1.8~\rm keV$, $Z = 0.014~Z_\odot$, and a normalization, $norm_{2,3}$, related to the gas density $n_{\text{H}} = 0.4~\rm cm^{-3}$ via Equation~(\ref{eq:norm}), and adopting, in this case, a volume equal to the volume of the shell. The resulting model was rescaled to express the emission in units of $\nu L_{\nu}$.

The panel (b) in Figure~\ref{fig:cloudymekal} presents the spectra obtained for the three models described above. The \texttt{Cloudy} thermal models with and without QSO photoionization are shown as blue and cyan lines, respectively, while the red line corresponds to the thermal emission from the \texttt{xsmekal} model. The gray transparent band highlights the energy range (0.5-2 keV observed) where extended X-ray emission is detected in the \textit{Chandra} data, and where we aim to identify differences between the models.
The panel (c) shows the ratio between the \texttt{Cloudy} and \texttt{xsmekal} thermal models without QSO photoionization, while the bottom panel displays the ratio between the \texttt{Cloudy} thermal models without and with the QSO radiation field. As shown by the similarity between the blue and cyan curves, whose ratio is shown in panel (d), the impact of QSO photoionization on the thermal emission of the hot gas is negligible in this case. 
Additionally, the comparison between the cyan and red curves, whose ratio is shown in panel (c), shows that, even under identical physical assumptions, the predicted emission varies only marginally between different modeling codes. These small differences are negligible for our purposes and do not affect our conclusion that QSO photoionization has no significant impact on the thermal emission of the hot gas. 

We also tested how the presence of the QSO affects the gas heating-cooling balance, and thus its cooling time, using time-dependent \texttt{Cloudy} simulations. We enabled non-equilibrium cooling (\texttt{set dynamics relax 3}) and tracked the thermal evolution over 300 iterations. A series of time steps, ranging from 10 to 100 Myr, was defined (using time commands), starting with an initial timestep of $10^4$ years. The simulation stops once the gas cools below $10^5$ K, allowing us to monitor and compare the cooling progression with and without QSO irradiation. We found that the cooling time to reach temperatures below $10^5~\rm K$ was $\sim 80$ Myr when the effect of QSO radiation on the thermal balance was neglected, and increased to $\sim 90$ Myr when this contribution was included.
This indicated that QSO photoionization could slow down the cooling rate \citep[e.g.,][]{Cantalupo10, Gnedin12}, although at temperatures $kT \lesssim 2~\rm keV$ the effect remained negligible, provided that the metallicity was low ($Z \lesssim 0.1~Z_\odot$).
We emphasize that these results do not imply that the gas will necessarily cool within 90 Myr, such as mechanical feedback or gravitational shocks, that are not included in the \textsc{Cloudy} calculations. The implications of the inferred cooling times are discussed further in Section~\ref{sec:cool}.

\subsection{Alternative scenarios for the extended soft X-ray emission}\label{sec:alternativemechanisms}

We developed analogous models to those of the previous section in which the hot thermal gas is replaced by cold/warm clouds photoionized by the QSO, exploring a grid of gas densities and metallicities. As expected, the emission from these photoionized clouds fails to account for the observed extended X-ray excess, even under the extreme assumption that the warm/cold phase occupies all the volume with filling factor $f_V=1$. 
A more detailed analysis of these models, including constraints derived from the \lya\ nebula, will be presented in a future paper, as this lies beyond the scope of the present work.\\

We tested a non-thermal IC scenario, fitting the X-ray excess with a power law, which results in unphysically steep photon indices of $\Gamma \sim 6$ \citep[see][]{Worrall16}, and found no evidence of powerful radio jets associated with this QSO (e.g., Sydney University Molonglo Sky Survey at 843 MHz; \citealt{Mauch03}, and 0.8 mJy/beam at 1.367 GHz and 0.78 mJy/beam at 887.5 MHz in the Rapid ASKAP Continuum Survey).

We also explored a scenario in which the extended X-ray emission originates from thermal Compton up-scattering of AGN seed photons by a hot electron population produced by an extended AGN wind, modeled using \texttt{compTT}. 
To reproduce the spectral shape of the extended X-ray emission, we assumed, as a conservative scenario, seed photons representative of the standard soft X-ray excess commonly observed in local AGN, typically peaking around $\sim 0.1~\rm keV$ rest-frame \citep{Crummy06,Done12}. 
We adopted an optical depth of $\tau = 0.01$, which is the upper limit derivable from our constraint on the electron column density in the X-ray quasar spectrum, corresponding to an electron density of $n_e = \tau / (l \, \sigma_{\rm T}) \sim 0.3~\rm cm^{-3}$, where $l \sim 15~\rm kpc$ and $\sigma_{\rm T} = 6.65 \times 10^{-25}~\rm cm^{2}$ is the Thomson cross-section. From spectral fitting, we obtained $\log(\texttt{norm}_{\texttt{compTT}}) = -2.77^{+0.71}_{-0.88}$ and $kT_e = 47^{+17}_{-11}~\rm keV$. 
Although this model can, in principle, reproduce the observed extended emission with a required  $L_{\text{seed}} \approx 2 \times 10^{46}~\rm erg~s^{-1}$, about a factor of two lower than the UV luminosity at 400$\, \AA$, we regard it as energetically implausible, as it requires the presence of a hot ($kT_e \sim 47~\rm keV$) and dense AGN wind that is capable of reaching distances of several tens of kiloparsecs from the nucleus. Although such AGN winds have been invoked by observations and numerical simulations \citep{FaucherGiguere12,Zubovas12,Weinberger17,Powell18}, these are typically confined within the galactic nucleus or the galaxy ISM, given their short cooling times. In addition, we do not see any signs of such fast and extended wind in the warm gas kinematics, as the \lya\ nebula appears kinematically quiet \citep{Borisova16} without detectable CIV emission. Similarly, we do not see any Broad Absorption Line in the quasar spectra. For all these reasons, we consider a Compton up-scattering origin of the detected X-ray emission as unlikely.

\subsection{Constraints from SZ non-detection}\label{sec:SZ}

The detection of extended X-ray emission suggests the presence of hot gas within a $\sim 3 \times 10^{13}~\rm M_{\odot}$ halo, potentially producing a thermal Sunyaev-Zeldovich (SZ) effect \citep{Sunyaev72}. However, an analysis of ALMA Band 3 data \citep{Pensabene24} reveals no clear evidence of an SZ signature in the current observations. To estimate the expected SZ significance ($\sigma_{\text{SZ}}$), we generate the SZ signature for each set of  $\beta$-model parameters, $kT$, and $n_{\text{e}}$ from the previous analyses, and inject the resulting models into jackknifed realizations of the available ALMA data \citep[see][for details on the jackknifing procedure]{DiMascolo23, vanMarrewijk25}. Our analysis indicates that the non-detection within the sensitivity limits of the current ALMA data is fully consistent with the hot gas properties as derived by the X-ray analysis.  
The posterior parameters from the MCMC analysis predict a marginal SZ signal with a significance of only $\simeq (2.6 \pm 0.4) \, \sigma_{\text{SZ}}$. Additionally, continuum emission from companion quasars and deviations from thermal pressure support could further suppress the SZ signal, potentially explaining the current non-detection. New ALMA Cycle 12 observations (Project ID: 2025.1.00107.S, priority grade B) have been awarded to our team. These data, once executed, will provide the sensitivity and angular resolution needed to further probe the thermal and non-thermal components of the ICM and to test more robustly the presence of an SZ signal in this system.

\subsection{Comparison with Spiderweb and simulated hot halos}\label{sec:compSpiderweb}

Studying the nascent hot phase of the CGM (or proto-ICM) in $z>2$ halos is difficult because it requires long integration times with current X-ray telescopes. Previous to this work, the only reported direct imaging detection of extended thermal ICM emission at $z>2$ was around the Spiderweb galaxy in \citetalias[$\approx 700~\rm ks$][]{Lepore24}, who reported residual extended emission after the subtraction of the X-ray Inverse Compton emission due to radio jets. 
For both Spiderweb and MQN01, the extended thermal emission appears to originate from the halo of a proto-BCG at the center of a forming (or proto-)cluster. 
\begin{figure}[t]
   \begin{center}
   \includegraphics[height=0.3\textheight,angle=0]{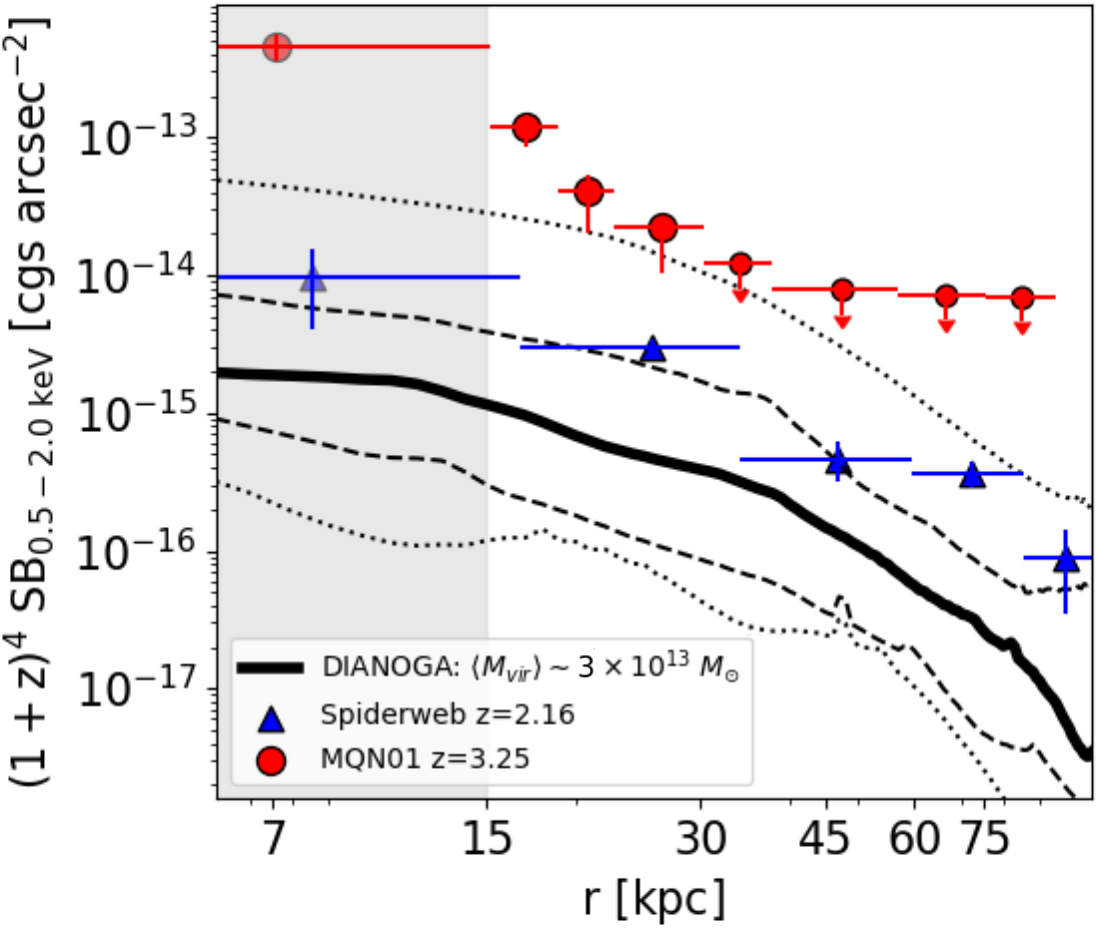}
\caption{Redshift-dimming corrected X-ray SB profiles in the observed 0.5-2 keV band ($\text{SB}_X$) for the extended thermal emission around QSO ID1 (red circles), obtained using an energy conversion factor (ECF) of $\simeq 4.6 \times 10^{-11}~\rm erg~cm^{-2}~cnts^{-1}$ based on the best-fit values of $kT = 1.8~\rm keV$ and $Z/Z_{\odot} = 0.014$, and for the Spiderweb Galaxy (blue triangles) \citepalias{Lepore24}. A correction factor of 0.536 was applied to the Spiderweb hot halo's $\text{SB}_X$ profile from \citetalias{Lepore24} to account for differences in the redshift-dependent intrinsic energy band. The black lines show the median (solid), 16th-84th percentile (dashed) and minimum-maximum (dotted) $\text{SB}_X$ profiles of hot halos at $z = 3$ from the \texttt{DIANOGA} simulations, selected to have an average mass consistent with the estimated virial mass of the MQN01 halo (i.e., $\sim 3 \times 10^{13}~M_{\odot}$).}
   \label{fig:profileSBMQNSpi}
   \end{center}
\end{figure}
Comparing luminosity, density, and morphology of these systems at $z=2.16$ and $z=3.25$ could provide some insights into the possible evolution of the hot CGM/proto-ICM with redshift and with their halo properties.  
Following the approach of \citetalias{Lepore24}, we computed the average SB of the extended X-ray emission in concentric annuli using:
\begin{equation} \label{eq:SBannulus}
	\text{SB}_{\text{Ann}} = ECF \times \frac{Net~Counts}{t_{\text{exp}} \times A_{\text{ann}}}\times \frac{ExpMap_{\text{max}}}{ExpMap_{\text{ann}}},
\end{equation}
where $A_{\text{ann}}$ is the area of the annulus, $t_{\text{exp}}$ is the total exposure time, and the ratio $ExpMap_{\text{max}}/ExpMap_{\text{ann}} \approx 1$ accounts for small variations of the \textit{Chandra} sensitivity within the relevant portion of the field of view. The energy conversion factor (ECF) depends on the temperature and metallicity parameters of the \texttt{xsmekal} model used to fit the spectra. The extended X-ray SB profile was obtained by subtracting the PSF-derived profile from the data. We estimated the SB profile in the 0.5-2 keV band (hereafter $\text{SB}_X (r)$) for the extended X-ray emission around QSO ID1 in the MQN01 protocluster. This used an ECF of $\sim 4.6 \times 10^{-11}~\rm erg~cm^{-2}~cnts^{-1}$, based on the best-fit temperature ($kT \sim 1.8~\rm keV$), and $0.014\, Z_{\odot}$ from our MCMC fit (see Section~\ref{sec:TB}). As indicated by the MCMC results, the PSF model was normalized to match the observed counts within the central 2$\arcsec$ (i.e., $\sim$15 kpc), under the assumption that $\sim 12 \, \%$ of the total emission originates from thermal processes (i.e., $f_{\rm th} = 0.12$; see Table~\ref{tab:flux_luminosity_unc}).

Figure~\ref{fig:profileSBMQNSpi} shows the redshift-dimming corrected $\text{SB}_X (r)$ of the extended X-ray emission in MQN01 (red) and Spiderweb (blue). 
The red dots, indicating the $\text{SB}_X (r)$ of the MQN01 halo, consist of three data points measured between 15 and 30 kpc (i.e., $2\arcsec$-$4\arcsec$; where the extended residual X-ray emission is detected following the subtraction of the nuclear emission), one point within 15 kpc ($<2\arcsec$; shaded grey region), which is extrapolated from the best-fitting $\beta$ model (see Section~\ref{sec:TB}), and upper limits at all radii above 30 kpc.  
The $\text{SB}_X$ profile of the X-ray extended halo in Spiderweb is taken from \citetalias{Lepore24}. To correct for different rest-frame bands due to redshift, we applied a 0.536 factor to the Spiderweb $\text{SB}_X(r)$ halo. This factor was derived assuming the best-fit $kT = 2~\rm keV$ \texttt{xsmekal} model reported in \citet{Tozzi22b}. 
The black lines show the $\text{SB}_X (r)$ profiles of simulated hot halos, where the solid, dashed, and dotted lines represent the median, 16th-84th percentile, and full range (minimum to maximum) of the $\text{SB}_X$ profiles, respectively.
These profiles were computed from a sample of 12 halos at $z=3$ extracted from the \texttt{DIANOGA} cosmological hydrodynamical simulations of galaxy clusters \citep{Esposito25}. This version of the \texttt{DIANOGA} simulations includes 14 target clusters at $z=0$, in the mass range $(0.2-3) \times 10^{15} ~\rm M_{\odot}$, simulated with \texttt{OpenGADGET-3} \citep[e.g.,][]{Groth23,Damiano24}. The 12 halos were selected in the mass range $(2-6) \times 10^{13} ~\rm M_{\odot}$ to have a median mass matching the virial mass of the MQN01 halo ($M_{\rm vir} \sim 3 \times 10^{13}~M_{\odot}$). The X-ray maps of the simulated halos were obtained with the post-processing tool SMAC \citep{Dolag05} in cubic regions of 500 kpc per side around the center of each halo, with 1 kpc resolution.

\begin{figure}[t]
   \begin{center}
   \includegraphics[height=0.29\textheight,angle=0]{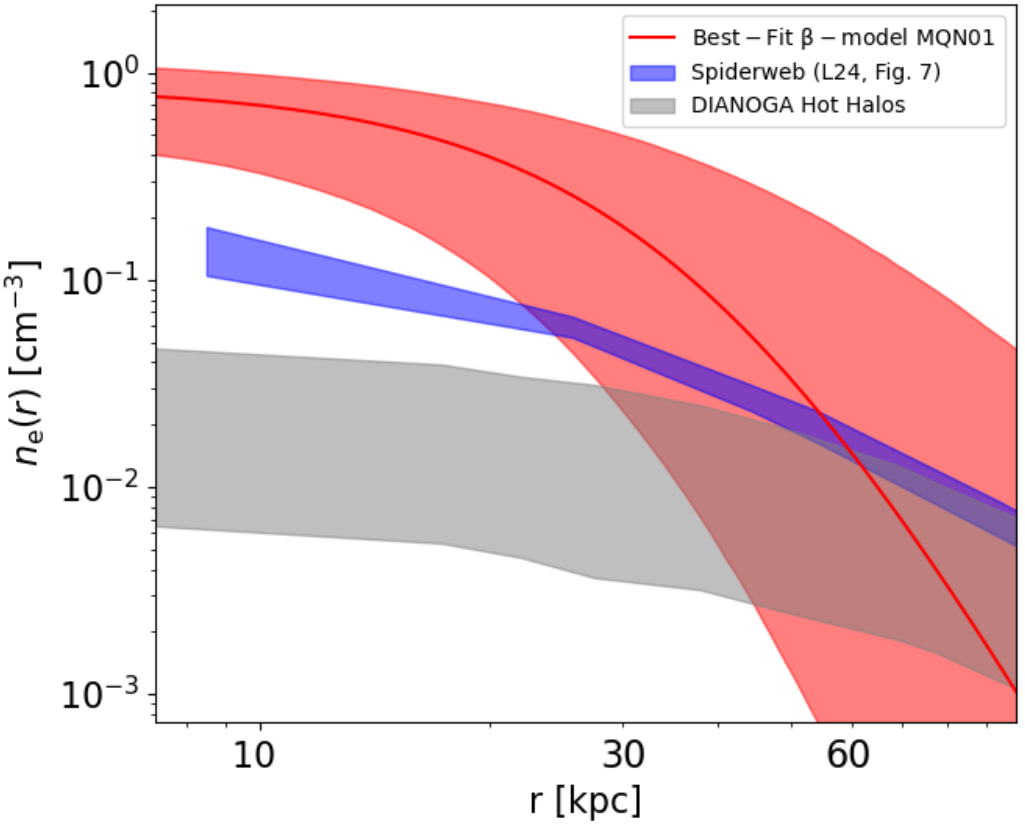}
\caption{Electron density profiles, $n_{\text{e}}(r)$, of the hot halos in MQN01 (red), Spiderweb (blue), and \texttt{DIANOGA} simulations (gray). The red curve shows the best-fit model for MQN01, with the shaded region indicating the 68\% confidence interval. The $n_{\text{e}}(r)$ profile of the Spiderweb halo is from \citetalias{Lepore24}, while the gray band represents the range of $n_{\text{e}}(r)$ profiles extracted from \texttt{DIANOGA} cosmological simulations of massive halos ($\sim$1-6 $\times 10^{13}~\rm M_{\odot}$) at similar redshift as MQN01 (i.e., $z = 3$).}
   \label{fig:profileDenSBMQNSpi}
   \end{center}
\end{figure}

While the $\text{SB}_X$ profiles of MQN01 and Spiderweb differ significantly in normalization, by a factor of 3-10 within 30 kpc from the QSOs, MQN01's profile is also noticeably steeper, consistent with a highly compact core. 
The $\text{SB}_X$ profile in MQN01 is consistent, for $r> 20~\rm kpc$, with the upper envelope of the $\text{SB}_X$ profiles from \texttt{DIANOGA} simulations. Similarly, the $\text{SB}_X$ profile of the Spiderweb halo closely follows the 84th percentile of the distribution from the \texttt{DIANOGA} hot halos. Within the limited statistics, simulations consistently predict fainter and shallower $\text{SB}_X$ profiles than observed. 
Part of the reason for this discrepancy may lie in the different sub-grid prescriptions adopted in the simulations. In particular, the \texttt{DIANOGA} runs use a relatively low density threshold for star formation ($0.13~\rm cm^{-3}$), which inhibits the formation of high-density gas regions, especially near halo centers. To test the impact of this, we compared the hot gas density profiles from \texttt{DIANOGA} with those of the most massive halos at $z=3$ in the \texttt{DaLya} simulations (priv. comm., \titu), which adopt a star formation threshold 100 times higher. For a consistent comparison, we applied the same temperature cut ($kT > 0.1~\rm keV$) as in the \texttt{DIANOGA} analysis to isolate the X-ray-emitting hot phase in both simulations. We found that the density of $kT \sim 0.1~\rm keV$ gas in \texttt{DaLya} is 5-10 times higher than in \texttt{DIANOGA}. While other factors, such as feedback modeling, numerical resolution, or selection effects in the observations, may also contribute to the discrepancy, the low star formation threshold in \texttt{DIANOGA} appears to suppress central gas densities and flatten the hot gas profiles (see Figure~\ref{fig:profileDenSBMQNSpi}), thus significantly reducing the X-ray emissivity of the hot CGM.\\

Figure~\ref{fig:profileDenSBMQNSpi} compares electron density profiles, $n_{\text{e}}(r)$, of hot halos in MQN01 (red line and black dots), Spiderweb (blue), and \texttt{DIANOGA} simulations (gray). The red curve shows the best-fit $n_{\text{e}}(r)$ for the MQN01 halo, with the shaded area indicating the 68\% confidence interval from Monte Carlo realizations on $n_{\text{e},0}$, $\beta$, and $r_{\text{core}}$. 
The $n_{\text{e}}(r)$ of the Spiderweb hot halo is from \citetalias{Lepore24}, while \texttt{DIANOGA} $n_{\text{e}}(r)$ are directly extracted from simulations. 

Between 15 and 30 kpc, the electron density of the MQN01 hot halo exceeds that of the Spiderweb halo by a factor of $4.3 \pm 1.0$ on average. At 15 kpc the ratio is $4.9 \pm 3.1$, decreasing to $3.5 \pm 4.5$ at 30 kpc (uncertainties propagated from the MQN01 $n_{\rm e}$ and the Spiderweb bounds). 

These SB$_X$ and $n_{\rm e}(r)$ differences broadly match expectations from cosmological evolution: in a self-similar scenario at fixed halo mass $n \propto (1+z)^3$ \citep{Kaiser86}, and since SB$_X \propto \int n_{\rm e} n_{\rm H}\Lambda(T,Z)\,dl$ one expects SB$_X \propto (1+z)^6$ for fixed $T$ and $Z$. For MQN01 ($z=3.25$) vs. Spiderweb ($z\sim 2.2$) this yields a factor $\sim 5.5$ in $\text{SB}_X$ and $\sim 2.34$ in $n_{\rm e}$. 
After accounting for this self-similar redshift scaling, the mean density ratio between 15-30 kpc range is reduced to $1.84 \pm 0.42$, with corrected point estimates of $2.08 \pm 1.32$ at 15 kpc and $1.48 \pm 1.91$ at 30 kpc (errors adjusted accordingly). The ratio at 30 kpc remains poorly constrained due to large uncertainties.
To account for unresolved gas inhomogeneities, we model the electron density at each radius as $n_{\text{e}}(r) = C^{1/2} \langle n_{\text{e}} \rangle$, where $C$ is the clumping factor \citep[e.g.,][]{Ettori13}. 
Reconciling the residual differences at 15 kpc would thus require modest clumping in MQN01 relative to the Spiderweb, with $C \simeq 4_{-3}^{+8}$. This range is consistent with typical clumping factors found in local ICM \citep[$ \leq 5$;][]{Nagai11,Vazza13,Eckert15,Mirakhor21}. However, the notably high central densities in MQN01 might also reflect additional effects, such as localized gas condensation or AGN-driven compression, which could enhance central gas densities beyond what is expected from redshift evolution alone.

\subsection{Pressure confinement of cold CGM clouds}\label{sec:pressureconfinement}

The hot gas pressure at 15 kpc and 30 kpc, that are the minimum and maximum radii at which we directly observe the extended X-ray emission based on our analysis, is $0.92 _{-0.63} ^{+1.24} ~ \rm keV~cm^{-3}$ and $0.30 _{-0.12} ^{+0.53} ~ \rm keV~cm^{-3}$, respectively. These are about one to two orders of magnitude higher than typical values observed for the ICM in local galaxy clusters \citep{Ettori09}, and about three/nine times higher than those obtained for the hot halo in Spiderweb \citetalias[i.e., $<0.1~\rm keV~cm^{-3}$][]{Lepore24} from independent measurements using ALMA SZ observation \citep{DiMascolo23}. Such a high thermal pressure is sufficient to confine the cold, \lya -emitting phase of the CGM. Assuming pressure equilibrium and a cold gas temperature of $T_{\rm{cold}} = 5 \times 10^4\rm ~K$, the implied temperature ratio $T_{\rm{hot}}/T_{\rm{cold}} \sim 400$ leads to a density contrast of the same factor. Therefore, for a hot gas density of $n_{\rm{hot}} = 0.1~\rm cm^{-3}$ at $\sim 50~\rm kpc$ (see Figure~\ref{fig:profileDenSBMQNSpi}), the warm gas can reach densities up to $n_{\rm{cold}} \sim 40~\rm cm^{-3}$, supporting a scenario in which the \lya\ nebula in MQN01 arises from a distribution of dense clumps pressure-confined by the surrounding hot halo \citep{Pezzulli19,Cantalupo19}, at least in the CGM inner regions. 
We stress, however, that our results (in particular the exceptionally high densities) may not be representative of the average \lya\ nebulae detected so far around luminous $z \sim 3$ quasars, as the latter do not generally reside in overdensities as large as that of MQN01 and their halo masses are expected to be, on average, smaller than what is derived for MQN01 \citep{Pezzulli19,deBeer23}.

\subsection{Comparison with local galaxy groups and clusters} \label{sec:ComparisonClusters}



\begin{figure}[t]
   \begin{center}
   \includegraphics[height=0.29\textheight,angle=0]{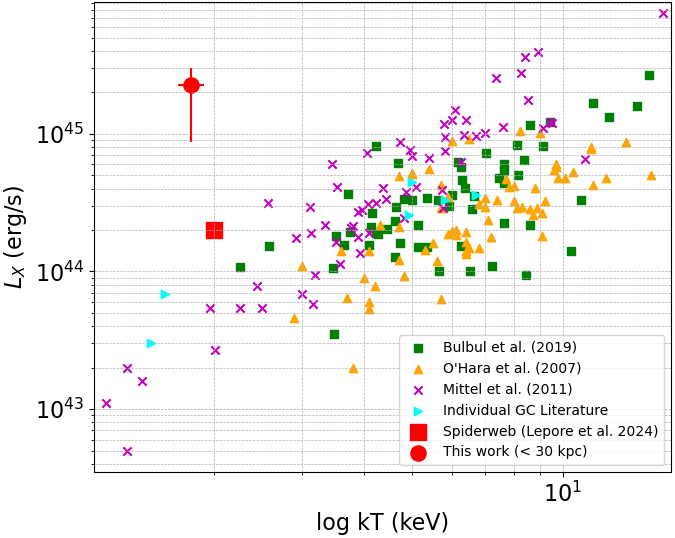}
\caption{Soft X-ray (0.5-2 keV) luminosity versus temperature $kT$ of the ICM in massive structures across redshift. Symbols show galaxy clusters from \citet{Bulbul19} (orange triangles), \citet{OHara07} (magenta crosses), \citet{Mittal11} (green squares), and high-redshift clusters from the literature \citep[cyan triangles;][]{Stanford01,Fabian01,Mathur03,Andreon11,Tozzi15,Mastromarino24}. The red square marks the Spiderweb protocluster hot X-ray halo \citepalias{Lepore24}, while red and orange circles show values for the X-ray halo in this work (MQN01), including and excluding the inner $<2\arcsec$ luminosity, respectively. The background density distributions represent the conservative MCMC posterior accounting for systematics (Section~\ref{sec:systematics}).}
   \label{fig:structureevolution}
   \end{center}
\end{figure}

Figure~\ref{fig:structureevolution} shows the relation between $L_{0.5-2~\rm keV}$ and $kT$ for a variety of massive structures, including low- and high-redshift systems ranging from galaxy groups to rich clusters. Notably, the red square represents the average temperature and X-ray luminosity of the extended X-ray emission in the Spiderweb protocluster ($kT=2.0_{-0.4}^{+0.7}~\rm keV$, $L_{0.5-2~\rm keV}= (2.0 \pm 0.5) \times 10^{44}~\rm erg~s^{-1}$). The Spiderweb point falls within the scatter of the group/cluster population, although the temperature is likely overestimated compared to the SZ-based value \citep[$kT_{SZ}=(0.7 \pm 0.3)~\rm keV$][]{DiMascolo23}. Adopting the lower SZ-based temperature would shift the Spiderweb point even further above the typical relation. 
The extended X-ray emission analyzed here at $z=3.25$ (red circle) shows a notably higher 0.5-2 keV luminosity than other structures with comparable temperatures, even after accounting for uncertainties on the $f_{th}$ value, placing MQN01 well outside the typical distribution of hot gas in groups and clusters.  

One possible explanation for the observed high luminosity could be related to redshift evolution. In particular, the self-similar evolution model \cite[see][where $L_X \propto E(z)$ as first approximation]{Maughan12} predicts luminosities of $\sim 6.3 \times 10^{43}~\rm erg~s^{-1}$ for Spiderweb and $\sim 4.1 \times 10^{44}~\rm erg~s^{-1}$ for MQN01. This provides a baseline to compare with local groups and clusters, under the assumption that the self-similar model applies. However, we note that this evolution is formally derived for bolometric luminosities, whereas our comparison is carried out in the soft band. Since the fraction of the bolometric luminosity falling in the soft band depends on the plasma temperature, applying the self-similar scaling directly to soft-band luminosities can overestimate the true redshift evolution, particularly for low-temperature systems. Even accounting for this, the MQN01 point remains significantly above the local $L_X$-$T$ population, while the Spiderweb point appears broadly consistent with it. 

Within the framework of thermal bremsstrahlung and line emission, this is not surprising, since we are probably observing a distinct and earlier phase of the ICM, which is characterized by different physical conditions. This pronounced deviation suggests that the enhanced luminosity of MQN01's diffuse X-ray gas may be due to higher gas densities at fixed mass (i.e., greater compactness), boosting emissivity. This compactness likely results from its higher redshift combined with a relatively steep gas density profile, supported by the $\beta$ value from our MCMC analysis (Table~\ref{tab:parMCMC_singleZ}), which exceeds typical local ICM values of 0.1-1 \citep{Mohr99,Dong10,Conte11,Paggi21,Xue00,Wise04,Mirakhor22}.

\subsection{Thermal instability and hot halo condensation}\label{sec:cool}

In this work, we have presented the first detection of extended hot ($T \gtrsim 10^7 K$) gas in a massive halo at $z>3$. Cosmological theory \citep[e.g.,][]{Dekel06} predicts that such structures should form as a consequence of the virialization of gas infalling onto dark matter haloes from the IGM and being shock-heated as a result of the collision of multiple infalling gas streams. The early cosmic time of our observation, along with with the peculiarly compact structure (see Section~\ref{sec:compSpiderweb}), leads us to speculate that we may be witnessing the initial stages of the formation and virialization of a hot halo, possibly leading, through further gas accretion and the expansion of the shock front to larger radii, to the more extended hot gas structures typically observed in the nearby Universe.

At the same time, the extreme luminosity of MQN01 (see Figure~\ref{fig:structureevolution}), indicative of large radiative losses, also raises the question of whether some of the recently virialized gas may already be undergoing substantial cooling, possibly contributing to the fueling of star formation and AGN activity in the central galaxy. A similar interpretation has been discussed by \citetalias{Lepore24} in the context of the hot halo of Spiderweb at $z \simeq 2.16$, based on the inferred low temperatures ($T \sim 0.3$-$0.9~\rm keV$) and short cooling times ($<100 ~\rm Myr$)\footnote{Shorter than those typically observed in cluster cores at low redshift \citep[$\approx 10^3~\rm Myr$;][]{Edge92,Hudson10}}. In addition, \citetalias{Lepore24} report a nominal isobaric mass deposition rate of $\dot{M}_{\text{cool}} \approx 0.25$-$1 \times 10^3~\rm M_{\odot}~yr^{-1}$, which exceeds typical cooling rates inferred for low-redshift ICMs \citep{McDonald18}, although it does not account for heating or feedback processes.

Similarly, applying the same steady-state, isobaric cooling-flow formalism to MQN01 \citep[Equation (2)][]{Fabian94}, and using $L_{\text{cool}} \approx L_{\rm 0.5-2\,keV}  ^{15-23 \rm~kpc} = 5.8_{-1.4}^{+2.1} \times 10^{44}~\rm erg~s^{-1}$ \citep[as in][]{Jones84}, would result in a mass deposition rate of $\dot{M}_{\text{cool}} \simeq 1.3_{-1.1}^{+1.7} \times 10^3~\rm M_{\odot}~yr^{-1}$ in this central region. This estimate assumes that the entire X-ray-emitting gas cools efficiently and steadily, and should be interpreted as an upper limit. We emphasise that these estimates ignore any source of heating or mechanical energy injection (e.g., AGN feedback, turbulence), which are thought to largely compensate for the cooling, so that the nominal mass deposition rates can be overestimated by factors 10-100 \citep[e.g.,][]{Fabian94,Gaspari13}. Moreover, a sustained mass accretion rate of $\simeq 10^3~\rm M_{\odot}~yr^{-1}$ could at most occur episodically. Indeed, if maintained for 1 Gyr, it would lead to an accumulation of $10^{12}~\rm M_{\odot}$ of cold gas, exceeding that observed in any known galaxy. 

To assess whether a less extreme scenario of localized thermal 
condensation is likely in our hot halo in MQN01, we compared the radiative cooling time, $t_{\rm cool}$, to the dynamical free-fall time, $t_{\rm ff}$, at $r=$15 and 30 kpc. The ratio $t_{\rm cool}/t_{\rm ff}$ is a widely used diagnostic for precipitation/condensation in hot halos \citep[e.g.][]{Voit15,Choudhury16,Voit17,Stern21,Donahue22}. 
Using the 'classical' formula (as shown by \citealt{Donahue22}) and a representative cooling function $\Lambda(T,Z) \simeq 10^{-23}  \rm erg~cm^{3}~s^{-1}$, we found $t_{\rm cool}(15\,{\rm kpc})=55 _{-14} ^{+24}$ Myr and $t_{\rm cool}(30\,{\rm kpc})=181 _{-83} ^{+268}$ Myr. 
Estimating $t_{\rm ff}$ from the local gravitational field \citep[e.g.,][]{McCourt12,Wibking25}, which we computed assuming a NFW potential with a concencentration $c\simeq 3.5 \pm 0.5$ \citep[as suggested by][for a $10^{13}~\rm M_{\odot}$ halo at $z \approx 3$]{Correa15}, we obtained $t_{\rm ff} (15\,{\rm kpc})=36 _{-4} ^{+5}$ Myr and $t_{\rm ff} (30\,{\rm kpc})=58 _{-5} ^{+8}$ Myr. These yield ratios $t_{\rm cool}/t_{\rm ff} (15\,{\rm kpc})=1.9 _{-0.9} ^{+1.9}$ and $t_{\rm cool}/t_{\rm ff} (30\,{\rm kpc})=3.1 _{-1.4} ^{+4.6}$. All fall within the commonly cited precipitation threshold, $1 < t_{\rm cool}/t_{\rm ff} < 10$, under which local thermal instabilities can seed multiphase condensation in roughly hydrostatic, feedback-regulated atmospheres \citep{Donahue22}. 

We note that, despite its common usage in the literature, the free-fall time $t_{\rm ff}$ is not strictly the most relevant timescale for local thermal instabilities. A more direct indicator of potential condensation is the Brunt-V\"ais\"al\"a time $t_{\text{BV}}$, which is the timescale for the reaction of buoyancy against thermal instability. 
While $t_{\text{ff}}$ and $t_{\text{BV}}$ are often comparable in compact halo cores, $t_{\text{BV}}$ remains a more precise diagnostic because it is defined locally through perturbation analysis \citep[e.g.,][]{Nipoti14,Wibking25}. To compute $t_{\text{BV}}$, we used Equation (D3) in Appendix~\ref{app:tBV}, again assuming the gravitational field $g$ associated with the NFW potential. We found $t_{\rm cool}/t_{\text{BV}}(15\,{\rm kpc})=1.3 _{-0.5} ^{+1.0}$,and $t_{\rm cool}/t_{\text{BV}}(30\,{\rm kpc})=4.5 _{-2.4} ^{+8.7}$, which are still in the range $1 < t_{\rm cool}/t_{\rm BV} < 10$. 

This does not imply a global cooling catastrophe, but it is consistent with intermittent condensation in portions of the hot halo that can feed cold inflows and/or SMBH fueling.  
Note, however, that classic analyses demonstrate that buoyancy in atmospheres with positive entropy gradients can suppress linear thermal instability \citep{Binney09}. Recent simulations \citep[see][]{Wibking25} further suggest that whether condensation occurs depends on turbulence, mixing, and the geometry of heating. Some systems with $t_{\rm cool}/t_{\rm ff}\lesssim10$ can remain stable if turbulent support is weak or heating is centrally concentrated \citep[see discussions in][]{Donahue22}. Consequently, our results should be interpreted as strong plausibility for localized condensation, contingent on the local balance of turbulence, mixing, and feedback.

To further probe the dynamical state of the hot gas halo, we examined the balance between the pressure gradient force and the gravitational acceleration derived from the NFW dark matter profile. Using the MCMC posterior parameters of $kT$, $r_{\text{core}}$, and $M_{\text{vir}}$ with their associated uncertainties, we computed the ratio of the hydrostatic acceleration from the pressure gradient, $1/ \rho (dP/dr)$, to the gravitational acceleration, $g(r)$, at radii of 15 and 30 kpc. Our Monte Carlo analysis, sampling the uncertainties, yields ratios of $\simeq 0.7_{-0.5} ^{+1.4}$ at 15 kpc and $\simeq 1.3 _{-0.8} ^{+1.7}$ at 30 kpc, consistent with hydrostatic equilibrium, although significant deviations are allowed within the credible intervals. These results suggest that, while the hot gas is roughly in hydrostatic balance, there may be local departures indicative of dynamical processes such as inflows, outflows, or turbulence. Such deviations can enhance thermal instability and promote condensation, consistent with the $1 < t_{\text{cool}}/t_{\text{ff}} < 10$ found above. Therefore, the combined diagnostics support a scenario in which the hot halo is marginally stable, possibly allowing localized cold gas condensation and potentially fueling galaxy growth and AGN activity.

\section{Summary and Conclusions}

In this paper, we report a significant (at least $8 \sigma$ between 15-30 kpc) detection of 0.5-2 keV ($\sim$2-8 keV rest-frame) X-ray emission, extended out to $\sim 30~\rm kpc$, around the brightest QSO (ID1) within the MQN01 protocluster at $z=3.25$ (Section~\ref{sec:results}). This QSO is also surrounded by a giant ($>200~\rm kpc$) \lya\ nebula \citep[\#1 in][]{Borisova16}. The extended X-ray emission exhibits an isotropic morphology and appears spatially correlated with the \lya\ nebula (Section~\ref{sec:morphology}). 

The steepness of the spectrum and the isotropic morphology of the extended X-ray emission suggest a thermal origin, with the emission powered by bremsstrahlung and collisional excitation mechanisms. Alternative emission mechanisms, such as photoionization of cold or hot gas clouds by the AGN, inverse Compton scattering, or Compton up-scattering, fail to fully or even partially reproduce the observed extended X-ray excess, as discussed in Section~\ref{sec:alternativemechanisms}.
Therefore, this likely represents the first evidence of thermal emission from proto-ICM (or hot CGM) at $z > 3$.

Assuming the QSO halo is virialized and its X-ray emission is characterized by a $\beta$-model, we performed a joint spatial and spectral MCMC analysis. The median posterior values imply the following (Section~\ref{sec:TB}):
\begin{itemize}
    \item A temperature of $\approx 1.8~\rm keV$ corresponding to a virial halo mass of $M_{\mathrm{vir}} \simeq (3 \pm 1) \times 10^{13}~M_{\odot}$.
    \item Hot gas within 30 kpc exhibits electron densities ranging from $0.9$ to $0.2~\rm cm^{-3}$ and a 0.5-2 keV X-ray luminosity of $L_{\text{0.5-2 keV}} = 2.25_{-1.38} ^{+0.77} \times 10^{45} ~\rm erg~s^{-1}$.
    \item The spatial fit yields $\beta$-model parameters of core radius $r_{\text{core}} \simeq 36_{-13}^{+16}~\rm kpc$ and $\beta \simeq 2.0_{-0.8}^{+1.4}$. The latter is significantly higher than the typical $\beta = 0.1$-$1$ observed in local galaxy groups and clusters, indicating a steeper gas density profile.
    \item The analysis implies that the hot gas in the halo of MQN01 contains a substantial fraction $f_\textrm{hot} = 56_{-20}^{+65} \%$ of the baryons that are theoretically associated to the halo (i.e.\ the baryonic mass that one would expect assuming a cosmic baryon fraction $\Omega_b/\Omega_m \simeq 0.15)$.
\end{itemize} 
Despite the large uncertainty, the last result, united to the fact that some baryonic mass must also be stored in the colder (Ly$\alpha$-emitting) phase of the CGM, suggests that the halo of MQN01 contains a substantial fraction of its theoretical baryon budget within the virial radius. We plan to better quantify the total baryon budget of MQN01 in a future combined analysis of both the X-ray and Ly$\alpha$ emission.

Based on our results, the currently available ALMA data are not expected to allow a clear detection of any thermal Sunyaev-Zeldovich (SZ) signal, as the expected significance is only $(2.6 \pm 0.4) \, \sigma_{\text{SZ}}$, even assuming perfect subtraction of contaminating sources (Section~\ref{sec:SZ}). However, upcoming tailored ALMA observations will be able to confirm (or challenge) our findings and add further constraints on the spatial distribution and thermal state of the hot gas in MQN01.

MQN01's hot halo is notably brighter and denser than that in the Spiderweb protocluster at $z=2.16$ \citepalias{Lepore24}, with X-ray surface brightness and electron density profiles exceeding those of Spiderweb by factors of 3-10 and 3-5 times within 30 kpc. These differences likely reflect cosmic evolution in density with redshift and potentially increased gas clumpiness. Compared to cosmological simulations (\texttt{DIANOGA}), MQN01's halo represents the luminous extreme of predicted hot halos, with higher densities possibly linked to a too low density threshold of star formation assumed in the simulations, which prevents dense gas from remaining in the X-ray emitting phase (Section~\ref{sec:compSpiderweb}).

On the other hand, the total X-ray luminosity within 30 kpc ($L_{0.5-2 ~\rm keV} \approx 2.3 \times 10^{45}\rm ~erg ~s^{-1}$) of the MQN01 halo is unusually high for its temperature, placing it well above the $L_X$-$kT$ relation for local groups and clusters. Even after correcting for self-similar evolution to $z=0$, MQN01 remains an outlier, reflecting a uniquely compact, dense gas distribution during an early intracluster medium formation phase. This enhanced emissivity likely arises from a steep density profile and concentrated gas within its dark matter halo, setting MQN01 apart from typical systems across cosmic time (Section~\ref{sec:ComparisonClusters}).

At roughly 15 kpc, MQN01's hot gas exhibits low temperature ($< 2~\rm keV$), short cooling times ($\approx 55~\rm Myr$), and relatively low entropies ($K \simeq 3~\rm keV~cm^{-2}$), and a relatively low cooling time to free-fall (or to Brunt-V\"ais\"al\"a time) time ratio $t_{\rm cool}/t_{\rm ff} \simeq 1.9_{-0.9}^{+1.9}$ ($t_{\rm cool}/t_{\rm BV} \simeq 1.3_{-0.5}^{+1.0}$), consistent with localized cold gas condensation according to some instability criteria proposed in the literature (Section~\ref{sec:cool}). On the other hand, a classical, large-scale cooling flow is deemed unlikely, but we report that the estimated mass deposition rate, assuming steady-state isobaric cooling, would be $\dot{M}_{\rm cool} \simeq 1.3_{-1.1}^{+1.7} \times 10^3~\rm M_{\odot}~yr^{-1}$, an extreme upper limit given likely heating and non-radiative effects (Section~\ref{sec:cool}).

The hot gas pressure at 15 kpc and 30 kpc ($\sim 0.30_{-0.12}^{+0.53}$ and $\sim 0.92_{-0.63}^{+1.24}~\rm keV~cm^{-3}$) are exceptionally high, one to two orders of magnitude above local cluster values and three to nine times higher than in the Spiderweb protocluster. Such pressure can efficiently pressure-confine denser \lya-emitting clouds, supporting a scenario for which the extended \lya\ nebula MQN01 could originate from colder clumps embedded within a hot, high-pressure medium.\\

Multi-wavelength follow-up observations with next-generation X-ray telescopes, such as the Advanced X-ray Imaging Satellite (AXIS) and NewAthena (Advanced Telescope for High ENergy Astrophysics), will be crucial to better characterize the physical properties of the extended gas distribution within the MQN01 Cosmic Web node. These will be enabled by their expected high spatial resolutions ($<1.5 \arcsec$ and $<3 \arcsec$) and spectral resolutions ($< 70 ~\rm eV$ and $< 1.5~\rm eV$ at 1 keV) in the soft X-ray band (0.3-10 keV and 0.2-12 keV) as reported in \cite{AXIS23} and \cite{Athena23}, respectively.
ALMA observations to detect the expected SZ signal on small scales are scheduled in Cycle 12 (Proposal ID 2025.1.01488.S). Very Long Baseline Interferometry (VLBI) radio observations may also prove valuable for exploring potential jet activity.
These investigations will be key to either confirming or challenging the thermal emission scenario proposed in this work, and to further constraining the physical properties of the nascent hot CGM/proto-ICM in this exceptional system.

\begin{acknowledgements}
This project was supported by the European Research Council (ERC) Consolidator Grant 864361 (CosmicWeb). FF acknowledges support by HORIZON2020: AHEAD2020-Grant Agreement n. 871158. PT acknowledges support from the Next Generation European Union PRIN 2022 20225E4SY5 - "From ProtoClusters to Clusters in one Gyr.". FV acknowledges support from the "INAF Ricerca Fondamentale 2023 Large GO" grant. LDM was supported by the French government, through the UCA\textsuperscript{J.E.D.I.} Investments in the Future project managed by the National Research Agency (ANR) with the reference number ANR-15-IDEX-01. RM acknowledges financial support from the ASI-INAF agreement n. 2022-14-HH.0. The authors also thank Alessandro Lupi, Silvano Molendi, Piero Rosati, and Ákos Bogdán for useful discussions. 
\end{acknowledgements}

\bibliographystyle{aa} 
\bibliography{andrea.bib}

\begin{thebibliography}{104}
\expandafter\ifx\csname natexlab\endcsname\relax\def\natexlab#1{#1}\fi

\bibitem[{{Andreon} \& {Huertas-Company}(2011)}]{Andreon11}
{Andreon}, S. \& {Huertas-Company}, M. 2011, A\&A, 526, A11

\bibitem[{{Arnaud}(1996)}]{Arnaud96}
{Arnaud}, K.~A. 1996, in Astronomical Society of the Pacific Conference Series, Vol. 101, Astronomical Data Analysis Software and Systems V, ed. G.~H. {Jacoby} \& J.~{Barnes}, 17

\bibitem[{{Asplund} {et~al.}(2009){Asplund}, {Grevesse}, {Sauval}, \& {Scott}}]{Asplund09}
{Asplund}, M., {Grevesse}, N., {Sauval}, A.~J., \& {Scott}, P. 2009, ARA\&A, 47, 481

\bibitem[{{Barret} {et~al.}(2023){Barret}, {Albouys}, {Herder}, {Piro}, {Cappi}, {Huovelin}, {Kelley}, {Mas-Hesse}, {Paltani}, {Rauw}, {Rozanska}, {Svoboda}, {Wilms}, {Yamasaki}, {Audard}, {Bandler}, {Barbera}, {Barcons}, {Bozzo}, {Ceballos}, {Charles}, {Costantini}, {Dauser}, {Decourchelle}, {Duband}, {Duval}, {Fiore}, {Gatti}, {Goldwurm}, {Hartog}, {Jackson}, {Jonker}, {Kilbourne}, {Korpela}, {Macculi}, {Mendez}, {Mitsuda}, {Molendi}, {Pajot}, {Pointecouteau}, {Porter}, {Pratt}, {Pr{\^e}le}, {Ravera}, {Sato}, {Schaye}, {Shinozaki}, {Skup}, {Soucek}, {Thibert}, {Vink}, {Webb}, {Chaoul}, {Raulin}, {Simionescu}, {Torrejon}, {Acero}, {Branduardi-Raymont}, {Ettori}, {Finoguenov}, {Grosso}, {Kaastra}, {Mazzotta}, {Miller}, {Miniutti}, {Nicastro}, {Sciortino}, {Yamaguchi}, {Beaumont}, {Cucchetti}, {D'Andrea}, {Eckart}, {Ferrando}, {Kammoun}, {Lotti}, {Mesnager}, {Natalucci}, {Peille}, {de Plaa}, {Ardellier}, {Argan}, {Bellouard}, {Carron}, {Cavazzuti}, {Fiorini}, {Khosropanah}, {Martin}, {Perry}, {Pinsard},
  {Pradines}, {Rigano}, {Roelfsema}, {Schwander}, {Torrioli}, {Ullom}, {Vera}, {Villegas}, {Zuchniak}, {Brachet}, {Cicero}, {Doriese}, {Durkin}, {Fioretti}, {Geoffray}, {Jacques}, {Kirsch}, {Smith}, {Adams}, {Gloaguen}, {Hoogeveen}, {van der Hulst}, {Kiviranta}, {van der Kuur}, {Ledot}, {van Leeuwen}, {van Loon}, {Lyautey}, {Parot}, {Sakai}, {van Weers}, {Abdoelkariem}, {Adam}, {Adami}, {Aicardi}, {Akamatsu}, {Alonso}, {Amato}, {Andr{\'e}}, {Angelinelli}, {Anon-Cancela}, {Anvar}, {Atienza}, {Attard}, {Auricchio}, {Balado}, {Bancel}, {Barusso}, {Bascu{\~n}an}, {Bernard}, {Berrocal}, {Blin}, {Bonino}, {Bonnet}, {Bonny}, {Boorman}, {Boreux}, {Bounab}, {Boutelier}, {Boyce}, {Brienza}, {Bruijn}, {Bulgarelli}, {Calarco}, {Callanan}, {Campello}, {Camus}, {Canourgues}, {Capobianco}, {Cardiel}, {Castellani}, {Cheatom}, {Chervenak}, {Chiarello}, {Clerc}, {Clerc}, {Cobo}, {Coeur-Joly}, {Coleiro}, {Colonges}, {Corcione}, {Coriat}, {Coynel}, {Cuttaia}, {D'Ai}, {D'anca}, {Dadina}, {Daniel}, {Dauner}, {DeNigris},
  {Dercksen}, {DiPirro}, {Doumayrou}, {Dubbeldam}, {Dupieux}, {Dupourqu{\'e}}, {Durand}, {Eckert}, {Eiriz}, {Ercolani}, {Etcheverry}, {Finkbeiner}, {Fiocchi}, {Fossecave}, {Franssen}, {Frericks}, {Gabici}, {Gant}, {Gao}, {Gastaldello}, \& {Genolet}}]{Athena23}
{Barret}, D., {Albouys}, V., {Herder}, J.-W.~d., {et~al.} 2023, ExA, 55, 373

\bibitem[{{Binney} {et~al.}(2009){Binney}, {Nipoti}, \& {Fraternali}}]{Binney09}
{Binney}, J., {Nipoti}, C., \& {Fraternali}, F. 2009, MNRAS, 397, 1804

\bibitem[{{Borisova} {et~al.}(2016){Borisova}, {Cantalupo}, {Lilly}, {Marino}, {Gallego}, {Bacon}, {Blaizot}, {Bouch{\'e}}, {Brinchmann}, {Carollo}, {Caruana}, {Finley}, {Herenz}, {Richard}, {Schaye}, {Straka}, {Turner}, {Urrutia}, {Verhamme}, \& {Wisotzki}}]{Borisova16}
{Borisova}, E., {Cantalupo}, S., {Lilly}, S.~J., {et~al.} 2016, \apj, 831, 39

\bibitem[{{Bulbul} {et~al.}(2019){Bulbul}, {Chiu}, {Mohr}, {McDonald}, {Benson}, {Bautz}, {Bayliss}, {Bleem}, {Brodwin}, {Bocquet}, {Capasso}, {Dietrich}, {Forman}, {Hlavacek-Larrondo}, {Holzapfel}, {Khullar}, {Klein}, {Kraft}, {Miller}, {Reichardt}, {Saro}, {Sharon}, {Stalder}, {Schrabback}, \& {Stanford}}]{Bulbul19}
{Bulbul}, E., {Chiu}, I.~N., {Mohr}, J.~J., {et~al.} 2019, ApJ, 871, 50

\bibitem[{{Cantalupo}(2010)}]{Cantalupo10}
{Cantalupo}, S. 2010, MNRAS, 403, L16

\bibitem[{{Cantalupo} {et~al.}(2019){Cantalupo}, {Pezzulli}, {Lilly}, {Marino}, {Gallego}, {Schaye}, {Bacon}, {Feltre}, {Kollatschny}, \& {Nanayakkara}}]{Cantalupo19}
{Cantalupo}, S., {Pezzulli}, G., {Lilly}, S.~J., {et~al.} 2019, \mnras, 483, 5188

\bibitem[{{Cavaliere} \& {Fusco-Femiano}(1976)}]{Cavaliere76}
{Cavaliere}, A. \& {Fusco-Femiano}, R. 1976, \aap, 49, 137

\bibitem[{{Choudhury} \& {Sharma}(2016)}]{Choudhury16}
{Choudhury}, P.~P. \& {Sharma}, P. 2016, MNRAS, 457, 2554

\bibitem[{{Conte} {et~al.}(2011){Conte}, {de Petris}, {Comis}, {Lamagna}, \& {de Gregori}}]{Conte11}
{Conte}, A., {de Petris}, M., {Comis}, B., {Lamagna}, L., \& {de Gregori}, S. 2011, A\&A, 532, A14

\bibitem[{{Correa} {et~al.}(2015){Correa}, {Wyithe}, {Schaye}, \& {Duffy}}]{Correa15}
{Correa}, C.~A., {Wyithe}, J. S.~B., {Schaye}, J., \& {Duffy}, A.~R. 2015, MNRAS, 452, 1217

\bibitem[{{Crummy} {et~al.}(2006){Crummy}, {Fabian}, {Gallo}, \& {Ross}}]{Crummy06}
{Crummy}, J., {Fabian}, A.~C., {Gallo}, L., \& {Ross}, R.~R. 2006, MNRAS, 365, 1067

\bibitem[{{Damiano} {et~al.}(2024){Damiano}, {Valentini}, {Borgani}, {Tornatore}, {Murante}, {Ragagnin}, {Ragone-Figueroa}, \& {Dolag}}]{Damiano24}
{Damiano}, A., {Valentini}, M., {Borgani}, S., {et~al.} 2024, arXiv, arXiv:2403.12600

\bibitem[{{de Beer} {et~al.}(2023){de Beer}, {Cantalupo}, {Travascio}, {Pezzulli}, {Galbiati}, {Fossati}, {Fumagalli}, {Lazeyras}, {Pensabene}, {Theuns}, \& {Wang}}]{deBeer23}
{de Beer}, S., {Cantalupo}, S., {Travascio}, A., {et~al.} 2023, MNRAS, 526, 1850

\bibitem[{{Dekel} \& {Birnboim}(2006)}]{Dekel06}
{Dekel}, A. \& {Birnboim}, Y. 2006, MNRAS, 368, 2

\bibitem[{{Di Mascolo} {et~al.}(2023){Di Mascolo}, {Saro}, {Mroczkowski}, {Borgani}, {Churazov}, {Rasia}, {Tozzi}, {Dannerbauer}, {Basu}, {Carilli}, {Ginolfi}, {Miley}, {Nonino}, {Pannella}, {Pentericci}, \& {Rizzo}}]{DiMascolo23}
{Di Mascolo}, L., {Saro}, A., {Mroczkowski}, T., {et~al.} 2023, \nat, 615, 809

\bibitem[{{Dolag} {et~al.}(2005){Dolag}, {Vazza}, {Brunetti}, \& {Tormen}}]{Dolag05}
{Dolag}, K., {Vazza}, F., {Brunetti}, G., \& {Tormen}, G. 2005, MNRAS, 364, 753

\bibitem[{{Donahue} \& {Voit}(2022)}]{Donahue22}
{Donahue}, M. \& {Voit}, G.~M. 2022, PhR, 973, 1

\bibitem[{{Done} {et~al.}(2012){Done}, {Davis}, {Jin}, {Blaes}, \& {Ward}}]{Done12}
{Done}, C., {Davis}, S.~W., {Jin}, C., {Blaes}, O., \& {Ward}, M. 2012, MNRAS, 420, 1848

\bibitem[{{Dong} {et~al.}(2010){Dong}, {Rasmussen}, \& {Mulchaey}}]{Dong10}
{Dong}, R., {Rasmussen}, J., \& {Mulchaey}, J.~S. 2010, ApJ, 712, 883

\bibitem[{{Eckert} {et~al.}(2013){Eckert}, {Molendi}, {Vazza}, {Ettori}, \& {Paltani}}]{Eckert13b}
{Eckert}, D., {Molendi}, S., {Vazza}, F., {Ettori}, S., \& {Paltani}, S. 2013, A\&A, 551, A22

\bibitem[{{Eckert} {et~al.}(2015){Eckert}, {Roncarelli}, {Ettori}, {Molendi}, {Vazza}, {Gastaldello}, \& {Rossetti}}]{Eckert15}
{Eckert}, D., {Roncarelli}, M., {Ettori}, S., {et~al.} 2015, MNRAS, 447, 2198

\bibitem[{{Edge} {et~al.}(1992){Edge}, {Stewart}, \& {Fabian}}]{Edge92}
{Edge}, A.~C., {Stewart}, G.~C., \& {Fabian}, A.~C. 1992, MNRAS, 258, 177

\bibitem[{{Ehlert} \& {Ulmer}(2009)}]{Ehlert09}
{Ehlert}, S. \& {Ulmer}, M.~P. 2009, \aap, 503, 35

\bibitem[{{Esposito} {et~al.}(2025){Esposito}, {Borgani}, {Strazzullo}, {Pannella}, {Granato}, {Ragone-Figueroa}, {Saro}, {Nonino}, \& {Valentini}}]{Esposito25}
{Esposito}, M., {Borgani}, S., {Strazzullo}, V., {et~al.} 2025, arXiv e-prints, arXiv:2503.01674

\bibitem[{{Ettori} {et~al.}(2013){Ettori}, {Donnarumma}, {Pointecouteau}, {Reiprich}, {Giodini}, {Lovisari}, \& {Schmidt}}]{Ettori13}
{Ettori}, S., {Donnarumma}, A., {Pointecouteau}, E., {et~al.} 2013, SSRv, 177, 119

\bibitem[{{Ettori} {et~al.}(2009){Ettori}, {Morandi}, {Tozzi}, {Balestra}, {Borgani}, {Rosati}, {Lovisari}, \& {Terenziani}}]{Ettori09}
{Ettori}, S., {Morandi}, A., {Tozzi}, P., {et~al.} 2009, A\&A, 501, 61

\bibitem[{{Fabian}(1994)}]{Fabian94}
{Fabian}, A.~C. 1994, ARA\&A, 32, 277

\bibitem[{{Fabian} {et~al.}(2001){Fabian}, {Crawford}, {Ettori}, \& {Sanders}}]{Fabian01}
{Fabian}, A.~C., {Crawford}, C.~S., {Ettori}, S., \& {Sanders}, J.~S. 2001, MNRAS, 322, L11

\bibitem[{{Faucher-Gigu{\`e}re} \& {Quataert}(2012)}]{FaucherGiguere12}
{Faucher-Gigu{\`e}re}, C.-A. \& {Quataert}, E. 2012, \mnras, 425, 605

\bibitem[{{Ferland} {et~al.}(1998){Ferland}, {Korista}, {Verner}, {Ferguson}, {Kingdon}, \& {Verner}}]{Ferland98}
{Ferland}, G.~J., {Korista}, K.~T., {Verner}, D.~A., {et~al.} 1998, \pasp, 110, 761

\bibitem[{{Foreman-Mackey} {et~al.}(2013){Foreman-Mackey}, {Hogg}, {Lang}, \& {Goodman}}]{Foreman-Mackey13}
{Foreman-Mackey}, D., {Hogg}, D.~W., {Lang}, D., \& {Goodman}, J. 2013, PASP, 125, 306

\bibitem[{{Freeman} {et~al.}(2001){Freeman}, {Doe}, \& {Siemiginowska}}]{Freeman01}
{Freeman}, P., {Doe}, S., \& {Siemiginowska}, A. 2001, in Society of Photo-Optical Instrumentation Engineers (SPIE) Conference Series, Vol. 4477, Astronomical Data Analysis, ed. J.-L. {Starck} \& F.~D. {Murtagh}, 76--87

\bibitem[{{Galbiati} {et~al.}(2024){Galbiati}, {Cantalupo}, {Steidel}, {Pensabene}, {Travascio}, {Wang}, {Fossati}, {Fumagalli}, {Rudie}, {Fresco}, {Lazeyras}, {Ledos}, \& {Quadri}}]{Galbiati24}
{Galbiati}, M., {Cantalupo}, S., {Steidel}, C., {et~al.} 2024, arXiv, arXiv:2410.03822

\bibitem[{{Gaspari} {et~al.}(2013){Gaspari}, {Ruszkowski}, \& {Oh}}]{Gaspari13}
{Gaspari}, M., {Ruszkowski}, M., \& {Oh}, S.~P. 2013, MNRAS, 432, 3401

\bibitem[{{Ge} {et~al.}(2018){Ge}, {Wang}, {Burchett}, {Tripp}, {Sun}, {Li}, {Gu}, \& {Ji}}]{Ge18}
{Ge}, C., {Wang}, Q.~D., {Burchett}, J.~N., {et~al.} 2018, MNRAS, 481, 4111

\bibitem[{{Gilli} {et~al.}(2007){Gilli}, {Comastri}, \& {Hasinger}}]{Gilli07}
{Gilli}, R., {Comastri}, A., \& {Hasinger}, G. 2007, \aap, 463, 79

\bibitem[{{Gnedin} \& {Hollon}(2012)}]{Gnedin12}
{Gnedin}, N.~Y. \& {Hollon}, N. 2012, ApJS, 202, 13

\bibitem[{{Gonzalez} {et~al.}(2007){Gonzalez}, {Zaritsky}, \& {Zabludoff}}]{Gonzalez07}
{Gonzalez}, A.~H., {Zaritsky}, D., \& {Zabludoff}, A.~I. 2007, ApJ, 666, 147

\bibitem[{Gradshteyn \& Ryzhik(2007)}]{Gradshteyn07}
Gradshteyn, I.~S. \& Ryzhik, I.~M. 2007, Table of integrals, series, and products, seventh edn. (Elsevier/Academic Press, Amsterdam), xlviii+1171, translated from the Russian, Translation edited and with a preface by Alan Jeffrey and Daniel Zwillinger, With one CD-ROM (Windows, Macintosh and UNIX)

\bibitem[{{Groth} {et~al.}(2023){Groth}, {Steinwandel}, {Valentini}, \& {Dolag}}]{Groth23}
{Groth}, F., {Steinwandel}, U.~P., {Valentini}, M., \& {Dolag}, K. 2023, MNRAS, 526, 616

\bibitem[{{Hinshaw} {et~al.}(2013){Hinshaw}, {Larson}, {Komatsu}, {Spergel}, {Bennett}, {Dunkley}, {Nolta}, {Halpern}, {Hill}, {Odegard}, {Page}, {Smith}, {Weiland}, {Gold}, {Jarosik}, {Kogut}, {Limon}, {Meyer}, {Tucker}, {Wollack}, \& {Wright}}]{Hinshaw13}
{Hinshaw}, G., {Larson}, D., {Komatsu}, E., {et~al.} 2013, ApJS, 208, 19

\bibitem[{{Hudson} {et~al.}(2010){Hudson}, {Mittal}, {Reiprich}, {Nulsen}, {Andernach}, \& {Sarazin}}]{Hudson10}
{Hudson}, D.~S., {Mittal}, R., {Reiprich}, T.~H., {et~al.} 2010, A\&A, 513, A37

\bibitem[{{Jones} \& {Forman}(1984)}]{Jones84}
{Jones}, C. \& {Forman}, W. 1984, ApJ, 276, 38

\bibitem[{{Kaiser}(1986)}]{Kaiser86}
{Kaiser}, N. 1986, MNRAS, 222, 323

\bibitem[{{Kooistra} {et~al.}(2022){Kooistra}, {Lee}, \& {Horowitz}}]{Kooistra22}
{Kooistra}, R., {Lee}, K.-G., \& {Horowitz}, B. 2022, \apj, 938, 123

\bibitem[{{Kravtsov} \& {Borgani}(2012)}]{Kravtsov12}
{Kravtsov}, A.~V. \& {Borgani}, S. 2012, ARA\&A, 50, 353

\bibitem[{{Lepore} {et~al.}(2024){Lepore}, {Di Mascolo}, {Tozzi}, {Churazov}, {Mroczkowski}, {Borgani}, {Carilli}, {Gaspari}, {Ginolfi}, {Liu}, {Pentericci}, {Rasia}, {Rosati}, {R{\"o}ttgering}, {Anderson}, {Dannerbauer}, {Miley}, \& {Norman}}]{Lepore24}
{Lepore}, M., {Di Mascolo}, L., {Tozzi}, P., {et~al.} 2024, \aap, 682, A186

\bibitem[{{Liedahl} {et~al.}(1995){Liedahl}, {Osterheld}, \& {Goldstein}}]{Liedahl95}
{Liedahl}, D.~A., {Osterheld}, A.~L., \& {Goldstein}, W.~H. 1995, \apjl, 438, L115

\bibitem[{{Mastromarino} {et~al.}(2024){Mastromarino}, {Oppizzi}, {De Luca}, {Bourdin}, \& {Mazzotta}}]{Mastromarino24}
{Mastromarino}, C., {Oppizzi}, F., {De Luca}, F., {Bourdin}, H., \& {Mazzotta}, P. 2024, A\&A, 688, A76

\bibitem[{{Mathur} \& {Williams}(2003)}]{Mathur03}
{Mathur}, S. \& {Williams}, R.~J. 2003, ApJL, 589, L1

\bibitem[{{Mauch} {et~al.}(2003){Mauch}, {Murphy}, {Buttery}, {Curran}, {Hunstead}, {Piestrzynski}, {Robertson}, \& {Sadler}}]{Mauch03}
{Mauch}, T., {Murphy}, T., {Buttery}, H.~J., {et~al.} 2003, MNRAS, 342, 1117

\bibitem[{{Maughan} {et~al.}(2012){Maughan}, {Giles}, {Randall}, {Jones}, \& {Forman}}]{Maughan12}
{Maughan}, B.~J., {Giles}, P.~A., {Randall}, S.~W., {Jones}, C., \& {Forman}, W.~R. 2012, MNRAS, 421, 1583

\bibitem[{{McCourt} {et~al.}(2012){McCourt}, {Sharma}, {Quataert}, \& {Parrish}}]{McCourt12}
{McCourt}, M., {Sharma}, P., {Quataert}, E., \& {Parrish}, I.~J. 2012, MNRAS, 419, 3319

\bibitem[{{McDonald} {et~al.}(2018){McDonald}, {Gaspari}, {McNamara}, \& {Tremblay}}]{McDonald18}
{McDonald}, M., {Gaspari}, M., {McNamara}, B.~R., \& {Tremblay}, G.~R. 2018, ApJ, 858, 45

\bibitem[{{Mewe} {et~al.}(1985){Mewe}, {Gronenschild}, \& {van den Oord}}]{Mewe85}
{Mewe}, R., {Gronenschild}, E.~H.~B.~M., \& {van den Oord}, G.~H.~J. 1985, \aaps, 62, 197

\bibitem[{{Mewe} {et~al.}(1986){Mewe}, {Lemen}, \& {van den Oord}}]{Mewe86}
{Mewe}, R., {Lemen}, J.~R., \& {van den Oord}, G.~H.~J. 1986, \aaps, 65, 511

\bibitem[{{Mirakhor} \& {Walker}(2021)}]{Mirakhor21}
{Mirakhor}, M.~S. \& {Walker}, S.~A. 2021, MNRAS, 506, 139

\bibitem[{{Mirakhor} {et~al.}(2022){Mirakhor}, {Walker}, \& {Runge}}]{Mirakhor22}
{Mirakhor}, M.~S., {Walker}, S.~A., \& {Runge}, J. 2022, MNRAS, 509, 1109

\bibitem[{{Mittal} {et~al.}(2011){Mittal}, {Hicks}, {Reiprich}, \& {Jaritz}}]{Mittal11}
{Mittal}, R., {Hicks}, A., {Reiprich}, T.~H., \& {Jaritz}, V. 2011, A\&A, 532, A133

\bibitem[{{Mohr} {et~al.}(1999){Mohr}, {Mathiesen}, \& {Evrard}}]{Mohr99}
{Mohr}, J.~J., {Mathiesen}, B., \& {Evrard}, A.~E. 1999, ApJ, 517, 627

\bibitem[{{Morandi} {et~al.}(2015){Morandi}, {Sun}, {Forman}, \& {Jones}}]{Morandi15}
{Morandi}, A., {Sun}, M., {Forman}, W., \& {Jones}, C. 2015, MNRAS, 450, 2261

\bibitem[{{Nagai} \& {Lau}(2011)}]{Nagai11}
{Nagai}, D. \& {Lau}, E.~T. 2011, ApJL, 731, L10

\bibitem[{{Navarro} {et~al.}(1996){Navarro}, {Frenk}, \& {White}}]{NFW96}
{Navarro}, J.~F., {Frenk}, C.~S., \& {White}, S. D.~M. 1996, ApJ, 462, 563

\bibitem[{{Nipoti} \& {Posti}(2014)}]{Nipoti14}
{Nipoti}, C. \& {Posti}, L. 2014, ApJ, 792, 21

\bibitem[{{O'Hara} {et~al.}(2007){O'Hara}, {Mohr}, \& {Sanderson}}]{OHara07}
{O'Hara}, T.~B., {Mohr}, J.~J., \& {Sanderson}, A.~J.~R. 2007, arXiv, arXiv:0710.5782

\bibitem[{{Paggi} {et~al.}(2021){Paggi}, {Massaro}, {Pe{\~n}a-Herazo}, {Missaglia}, {Ricci}, {Stuardi}, {Kraft}, {Tremblay}, {Baum}, \& {Wilkes}}]{Paggi21}
{Paggi}, A., {Massaro}, F., {Pe{\~n}a-Herazo}, H.~A., {et~al.} 2021, A\&A, 647, A79

\bibitem[{{Peebles}(1980)}]{Peebles80}
{Peebles}, P.~J.~E. 1980, {The large-scale structure of the universe}

\bibitem[{{Pensabene} {et~al.}(2024){Pensabene}, {Cantalupo}, {Cicone}, {Decarli}, {Galbiati}, {Ginolfi}, {de Beer}, {Fossati}, {Fumagalli}, {Lazeyras}, {Pezzulli}, {Travascio}, {Wang}, {Matthee}, \& {Maseda}}]{Pensabene24}
{Pensabene}, A., {Cantalupo}, S., {Cicone}, C., {et~al.} 2024, arXiv e-prints, arXiv:2401.04765

\bibitem[{{Pezzulli} \& {Cantalupo}(2019)}]{Pezzulli19}
{Pezzulli}, G. \& {Cantalupo}, S. 2019, \mnras, 486, 1489

\bibitem[{{Pezzulli} {et~al.}(2017){Pezzulli}, {Fraternali}, \& {Binney}}]{Pezzulli17}
{Pezzulli}, G., {Fraternali}, F., \& {Binney}, J. 2017, MNRAS, 467, 311

\bibitem[{{Powell} {et~al.}(2018){Powell}, {Husemann}, {Tremblay}, {Krumpe}, {Urrutia}, {Baum}, {Busch}, {Combes}, {Croom}, {Davis}, {Eckart}, {O'Dea}, {P{\'e}rez-Torres}, {Scharw{\"a}chter}, {Smirnova-Pinchukova}, \& {Urry}}]{Powell18}
{Powell}, M.~C., {Husemann}, B., {Tremblay}, G.~R., {et~al.} 2018, A\&A, 618, A27

\bibitem[{{Pratt} {et~al.}(2023){Pratt}, {Arnaud}, {Maughan}, \& {Melin}}]{Pratt23}
{Pratt}, G.~W., {Arnaud}, M., {Maughan}, B.~J., \& {Melin}, J.~B. 2023, A\&A, 669, C2

\bibitem[{{Reynolds} {et~al.}(2023){Reynolds}, {Kara}, {Mushotzky}, {Ptak}, {Koss}, {Williams}, {Allen}, {Bauer}, {Bautz}, {Bogadhee}, {Burdge}, {Cappelluti}, {Cenko}, {Chartas}, {Chan}, {Corrales}, {Daylan}, {Falcone}, {Foord}, {Grant}, {Habouzit}, {Haggard}, {Herrmann}, {Hodges-Kluck}, {Kargaltsev}, {King}, {Kounkel}, {Lopez}, {Marchesi}, {McDonald}, {Meyer}, {Miller}, {Nynka}, {Okajima}, {Pacucci}, {Russell}, {Safi-Harb}, {Strassun}, {Trindade Falc{\~a}o}, {Walker}, {Wilms}, {Yukita}, \& {Zhang}}]{AXIS23}
{Reynolds}, C.~S., {Kara}, E.~A., {Mushotzky}, R.~F., {et~al.} 2023, in Society of Photo-Optical Instrumentation Engineers (SPIE) Conference Series, Vol. 12678, UV, X-Ray, and Gamma-Ray Space Instrumentation for Astronomy XXIII, ed. O.~H. {Siegmund} \& K.~{Hoadley}, 126781E

\bibitem[{{Rosati} {et~al.}(2002){Rosati}, {Tozzi}, {Giacconi}, {Gilli}, {Hasinger}, {Kewley}, {Mainieri}, {Nonino}, {Norman}, {Szokoly}, {Wang}, {Zirm}, {Bergeron}, {Borgani}, {Gilmozzi}, {Grogin}, {Koekemoer}, {Schreier}, \& {Zheng}}]{Rosati02}
{Rosati}, P., {Tozzi}, P., {Giacconi}, R., {et~al.} 2002, \apj, 566, 667

\bibitem[{{Ruppin} {et~al.}(2021){Ruppin}, {McDonald}, {Bleem}, {Allen}, {Benson}, {Calzadilla}, {Khullar}, \& {Floyd}}]{Ruppin21}
{Ruppin}, F., {McDonald}, M., {Bleem}, L.~E., {et~al.} 2021, ApJ, 918, 43

\bibitem[{{Saro} {et~al.}(2009){Saro}, {Borgani}, {Tornatore}, {De Lucia}, {Dolag}, \& {Murante}}]{Saro09}
{Saro}, A., {Borgani}, S., {Tornatore}, L., {et~al.} 2009, MNRAS, 392, 795

\bibitem[{{Sato} {et~al.}(2000){Sato}, {Akimoto}, {Furuzawa}, {Tawara}, {Watanabe}, \& {Kumai}}]{Sato00}
{Sato}, S., {Akimoto}, F., {Furuzawa}, A., {et~al.} 2000, ApJL, 537, L73

\bibitem[{{Shimakawa} {et~al.}(2018){Shimakawa}, {Koyama}, {R{\"o}ttgering}, {Kodama}, {Hayashi}, {Hatch}, {Dannerbauer}, {Tanaka}, {Tadaki}, {Suzuki}, {Fukagawa}, {Cai}, \& {Kurk}}]{Shimakawa18}
{Shimakawa}, R., {Koyama}, Y., {R{\"o}ttgering}, H. J.~A., {et~al.} 2018, \mnras, 481, 5630

\bibitem[{{Siemiginowska} {et~al.}(2024){Siemiginowska}, {Burke}, {G{\"u}nther}, {Lee}, {McLaughlin}, {Principe}, {Cheer}, {Fruscione}, {Laurino}, {McDowell}, \& {Terrell}}]{Siemiginowska24}
{Siemiginowska}, A., {Burke}, D., {G{\"u}nther}, H.~M., {et~al.} 2024, \apjs, 274, 43

\bibitem[{{Smail} {et~al.}(2012){Smail}, {Blundell}, {Lehmer}, \& {Alexander}}]{Smail12}
{Smail}, I., {Blundell}, K.~M., {Lehmer}, B.~D., \& {Alexander}, D.~M. 2012, ApJ, 760, 132

\bibitem[{{Stanford} {et~al.}(2001){Stanford}, {Holden}, {Rosati}, {Tozzi}, {Borgani}, {Eisenhardt}, \& {Spinrad}}]{Stanford01}
{Stanford}, S.~A., {Holden}, B., {Rosati}, P., {et~al.} 2001, ApJ, 552, 504

\bibitem[{{Stern} {et~al.}(2021){Stern}, {Faucher-Gigu{\`e}re}, {Fielding}, {Quataert}, {Hafen}, {Gurvich}, {Ma}, {Byrne}, {El-Badry}, {Angl{\'e}s-Alc{\'a}zar}, {Chan}, {Feldmann}, {Kere{\v{s}}}, {Wetzel}, {Murray}, \& {Hopkins}}]{Stern21}
{Stern}, J., {Faucher-Gigu{\`e}re}, C.-A., {Fielding}, D., {et~al.} 2021, ApJ, 911, 88

\bibitem[{{Sunyaev} \& {Zeldovich}(1972)}]{Sunyaev72}
{Sunyaev}, R.~A. \& {Zeldovich}, Y.~B. 1972, CoASP, 4, 173

\bibitem[{{Tozzi} {et~al.}(2022){Tozzi}, {Gilli}, {Liu}, {Borgani}, {Lepore}, {Di Mascolo}, {Saro}, {Pentericci}, {Carilli}, {Miley}, {Mroczkowski}, {Pannella}, {Rasia}, {Rosati}, {Anderson}, {Calabr{\'o}}, {Churazov}, {Dannerbauer}, {Feruglio}, {Fiore}, {Gobat}, {Jin}, {Nonino}, {Norman}, \& {R{\"o}ttgering}}]{Tozzi22b}
{Tozzi}, P., {Gilli}, R., {Liu}, A., {et~al.} 2022, A\&A, 667, A134

\bibitem[{{Tozzi} {et~al.}(2015){Tozzi}, {Santos}, {Jee}, {Fassbender}, {Rosati}, {Nastasi}, {Forman}, {Sartoris}, {Borgani}, {Boehringer}, {Altieri}, {Pratt}, {Nonino}, \& {Jones}}]{Tozzi15}
{Tozzi}, P., {Santos}, J.~S., {Jee}, M.~J., {et~al.} 2015, \apj, 799, 93

\bibitem[{{Travascio} {et~al.}(2024){Travascio}, {Cantalupo}, {Tozzi}, {Vito}, {Paggi}, {Pezzulli}, {Elvis}, {Fabbiano}, {Fiore}, {Fossati}, {Fresco}, {Fumagalli}, {Galbiati}, {Lazeyras}, {Ledos}, {Pannella}, {Pensabene}, {Quadri}, \& {Wang}}]{Travascio24b}
{Travascio}, A., {Cantalupo}, S., {Tozzi}, P., {et~al.} 2024, arXiv, arXiv:2410.03933

\bibitem[{{van Marrewijk} {et~al.}(2024){van Marrewijk}, {Di Mascolo}, {Gill}, {Battaglia}, {Battistelli}, {Bond}, {Devlin}, {Doze}, {Dunkley}, {Knowles}, {Hincks}, {Hughes}, {Hilton}, {Moodley}, {Mroczkowski}, {Naess}, {Partridge}, {Popping}, {Sif{\'o}n}, {Staggs}, \& {Wollack}}]{Marrewijk24}
{van Marrewijk}, J., {Di Mascolo}, L., {Gill}, A.~S., {et~al.} 2024, \aap, 689, A41

\bibitem[{{van Marrewijk} {et~al.}(2025){van Marrewijk}, {Kaasinen}, {Popping}, {Di Mascolo}, {Mroczkowski}, {Boogaard}, {Valentino}, {Bakx}, \& {Yoon}}]{vanMarrewijk25}
{van Marrewijk}, J., {Kaasinen}, M., {Popping}, G., {et~al.} 2025, A\&A, 695, A204

\bibitem[{{Vazza} {et~al.}(2013){Vazza}, {Eckert}, {Simionescu}, {Br{\"u}ggen}, \& {Ettori}}]{Vazza13}
{Vazza}, F., {Eckert}, D., {Simionescu}, A., {Br{\"u}ggen}, M., \& {Ettori}, S. 2013, MNRAS, 429, 799

\bibitem[{{Vikhlinin} {et~al.}(2005){Vikhlinin}, {Markevitch}, {Murray}, {Jones}, {Forman}, \& {Van Speybroeck}}]{Vikhlinin05}
{Vikhlinin}, A., {Markevitch}, M., {Murray}, S.~S., {et~al.} 2005, ApJ, 628, 655

\bibitem[{{Voit} {et~al.}(2015){Voit}, {Donahue}, {Bryan}, \& {McDonald}}]{Voit15}
{Voit}, G.~M., {Donahue}, M., {Bryan}, G.~L., \& {McDonald}, M. 2015, Natur, 519, 203

\bibitem[{{Voit} {et~al.}(2017){Voit}, {Meece}, {Li}, {O'Shea}, {Bryan}, \& {Donahue}}]{Voit17}
{Voit}, G.~M., {Meece}, G., {Li}, Y., {et~al.} 2017, ApJ, 845, 80

\bibitem[{{Wang} \& {Abel}(2008)}]{Wang08}
{Wang}, P. \& {Abel}, T. 2008, \apj, 672, 752

\bibitem[{{Wang} {et~al.}(2016){Wang}, {Elbaz}, {Daddi}, {Finoguenov}, {Liu}, {Schreiber}, {Mart{\'\i}n}, {Strazzullo}, {Valentino}, {van der Burg}, {Zanella}, {Ciesla}, {Gobat}, {Le Brun}, {Pannella}, {Sargent}, {Shu}, {Tan}, {Cappelluti}, \& {Li}}]{WangElbaz16}
{Wang}, T., {Elbaz}, D., {Daddi}, E., {et~al.} 2016, \apj, 828, 56

\bibitem[{{Weinberger} {et~al.}(2017){Weinberger}, {Springel}, {Hernquist}, {Pillepich}, {Marinacci}, {Pakmor}, {Nelson}, {Genel}, {Vogelsberger}, {Naiman}, \& {Torrey}}]{Weinberger17}
{Weinberger}, R., {Springel}, V., {Hernquist}, L., {et~al.} 2017, MNRAS, 465, 3291

\bibitem[{{Wibking} {et~al.}(2025){Wibking}, {Voit}, \& {O'Shea}}]{Wibking25}
{Wibking}, B.~D., {Voit}, G.~M., \& {O'Shea}, B.~W. 2025, MNRAS, 537, 739

\bibitem[{{Wise} {et~al.}(2004){Wise}, {McNamara}, \& {Murray}}]{Wise04}
{Wise}, M.~W., {McNamara}, B.~R., \& {Murray}, S.~S. 2004, ApJ, 601, 184

\bibitem[{{Worrall} {et~al.}(2016){Worrall}, {Birkinshaw}, \& {Young}}]{Worrall16}
{Worrall}, D.~M., {Birkinshaw}, M., \& {Young}, A.~J. 2016, MNRAS, 458, 174

\bibitem[{{Xue} \& {Wu}(2000)}]{Xue00}
{Xue}, Y.-J. \& {Wu}, X.-P. 2000, MNRAS, 318, 715

\bibitem[{Yuan {et~al.}(2003)Yuan, Fabian, Celotti, \& Jonker}]{Yuan03}
Yuan, W., Fabian, A.~C., Celotti, A., \& Jonker, P.~G. 2003, Monthly Notices of the Royal Astronomical Society, 346, L7–L10

\bibitem[{{Zubovas} \& {King}(2012)}]{Zubovas12}
{Zubovas}, K. \& {King}, A. 2012, \apjl, 745, L34

\end{thebibliography}


\appendix 
\onecolumn

\section{Alignment of QSO ID1}\label{app:merged}

\begin{figure*}[t]
   \begin{center}
   \includegraphics[height=0.42\textheight,angle=0]{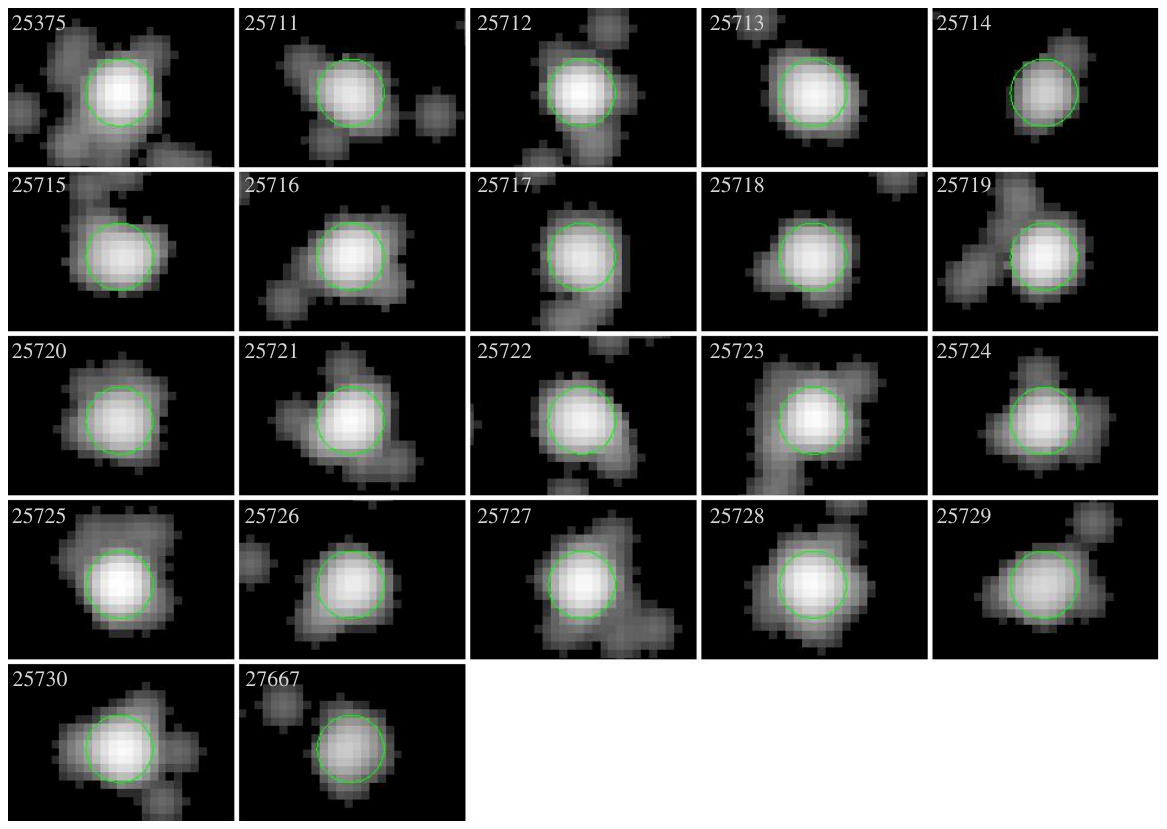}
   \caption{Images at the 0.5-2 keV energy band of all the ObsID (marked on the top-left in each panel), centered on the QSO ID1. The images are binned to one-fourth of the native pixel size and smoothed using a Gaussian kernel of 3 sub-pixels. The green 2$\arcsec$ circle indicates the QSO center, which was used for alignment across observations.}
   \label{fig:reproj}
   \end{center}
\end{figure*}

Since the primary aim of this paper is to investigate potential extended X-ray emission around QSOs, with a particular focus on QSO ID1 due to its high photon count, we developed a refined alignment and merging procedure for the individual ObsIDs. This method minimizes misalignment in the PSF of the ID1 source across different observations, allowing for a more reliable search for extended X-ray emission in the merged event file. Section~\ref{sec:align} contains a more detailed description of this method.
Figure~\ref{fig:reproj} presents the results of this alignment, showing the 0.5-2 keV images of the ObsIDs, centered on the QSO ID1, after re-projection and Gaussian smoothing. The green circles mark 2$\arcsec$-radius regions centered on a common reference coordinate corresponding to the QSO position. This alignment, along with additional tests discussed in the main text, confirms that the observed extended emission is not an artifact caused by ObsID misalignment or inaccuracies in PSF modeling.

\section{Lack of Extended X-ray Emission around the others AGN within MQN01} \label{app:nores}

\begin{figure*}[t]
   \begin{center}
   \includegraphics[height=0.45\textheight,angle=0]{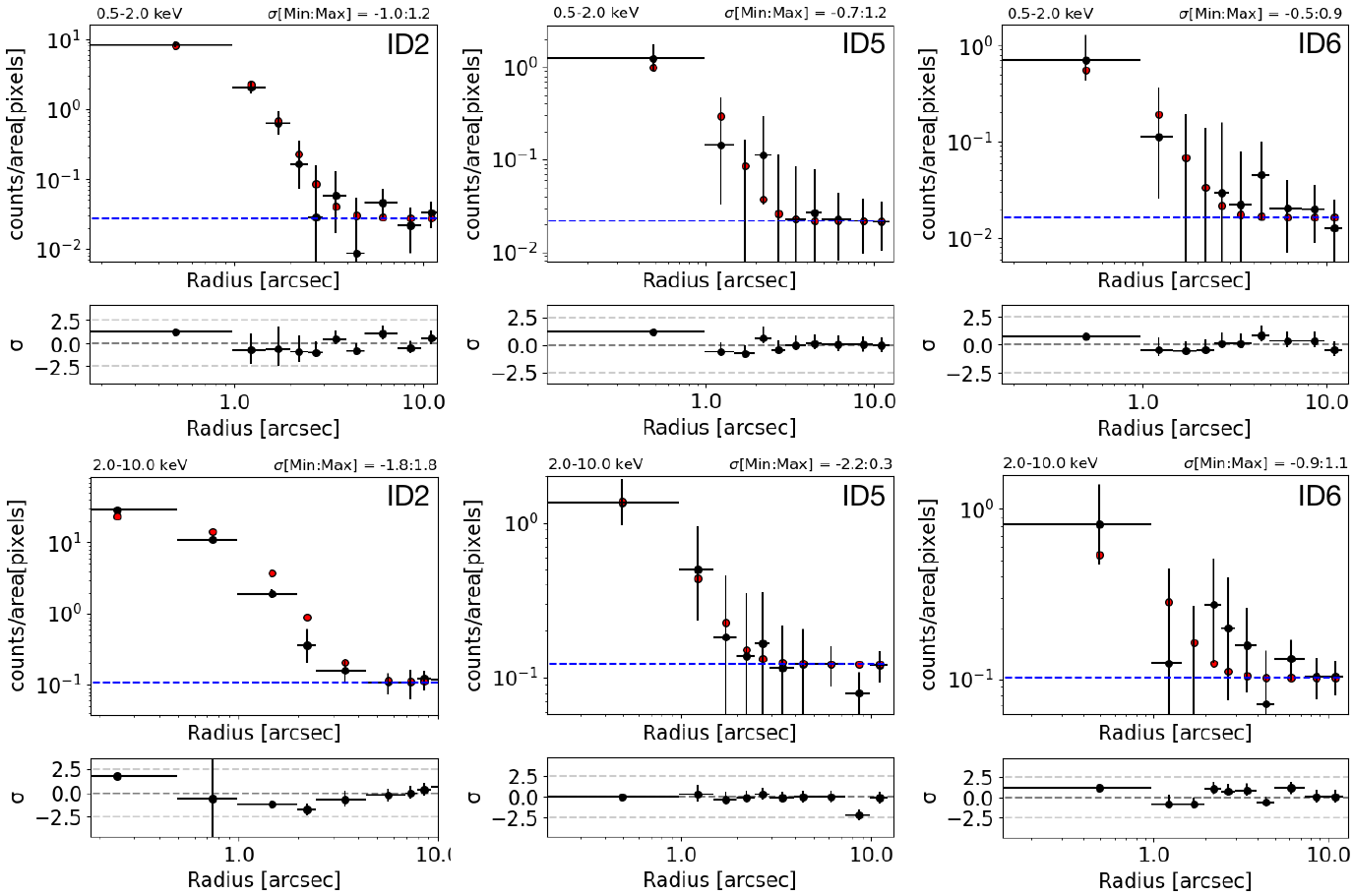}
   \caption{Radial profiles of surface counts and residuals, as shown in Figure~\ref{fig:rp}, centered on the AGNs QSO ID2, ID5, and ID6 \citep[from][]{Travascio24b}. The profiles are derived from observational data (black dots) and the simulated PSF (red dots). The top and bottom panels display the profiles in the 0.5-2 keV and 2-10 keV energy bands, respectively.}
   \label{fig:nores}
   \end{center}
\end{figure*}

\begin{figure}[t]
   \begin{center}
   \includegraphics[height=0.2\textheight,angle=0]{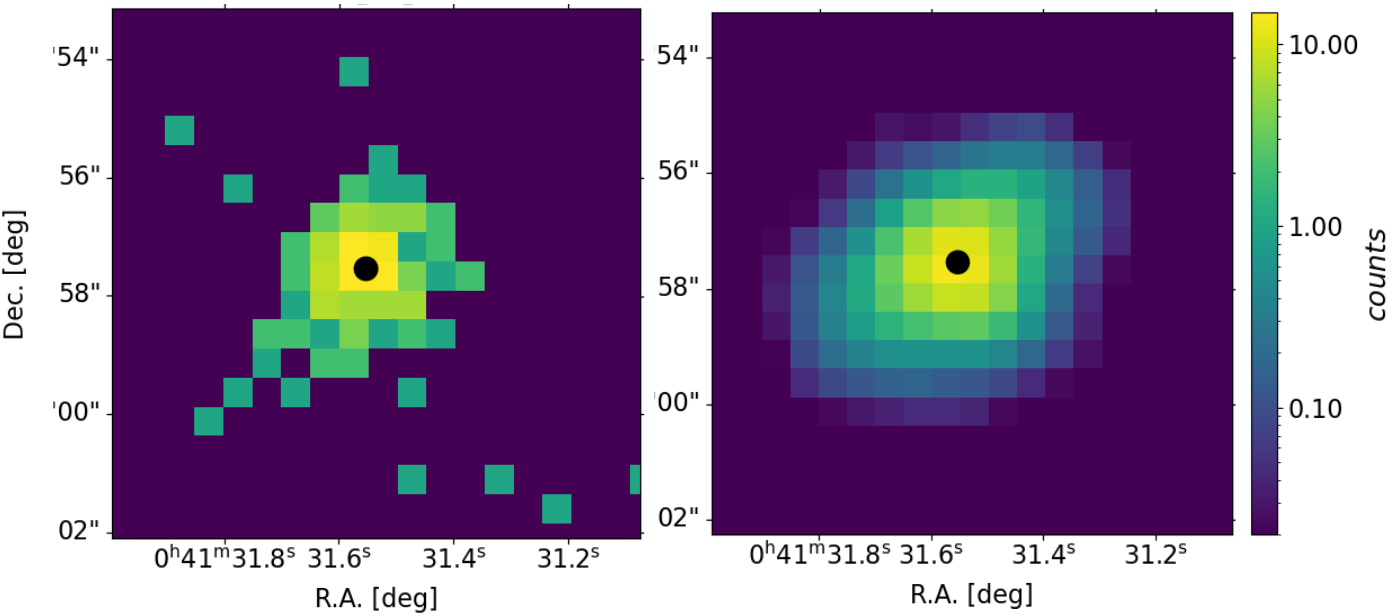}
   \caption{The same plot as in the left panel in Figure~\ref{fig:rp2} (see here for further details), for the second-brightest X-ray AGN, ID2.}
   \label{fig:imagesID2datapsf}
   \end{center}
\end{figure}

Figure~\ref{fig:nores} presents the radial profiles extracted from observational data (black points) compared to those derived from the simulated PSF (red points) for other X-ray AGNs reported in \cite{Travascio24b}. AGNs ID3 and ID4, located near the brightest QSO (ID1), were excluded due to their proximity, which complicates the identification of extended emission. As reported in Section~\ref{sec:detection}, no significant extended emission is detected around these AGNs, as the normalized PSF accurately reproduces the observed profiles within the uncertainties. Beyond 4$\arcsec$ (corresponding to 30 kpc at $z = 3.25$), the uncertainties are primarily dominated by the local background. This result further validates the reliability of our analysis and the accuracy of the generated PSFs in reproducing the observational data.

To further test this, we estimated the ratio of net counts in the 0.5-1.5 keV energy range (where the extended emission is detected) within the 2$\arcsec$-3$\arcsec$ annulus and the 2$\arcsec$-radius region centered on ID1. This was then compared to the ratios for other bright X-ray point sources in the \textit{Chandra} field of view (FoV), located as close as possible to ID1. Among these, the only source with a well-defined, non-elongated PSF, a high count rate, and a position near the aim point is the AGN ID2. 
To estimate the net counts, the background was derived from 100 large annuli around the sources, with their minimum and maximum radii chosen randomly within specific limits to minimize the influence of local background variations. By computing the median, along with the 25th and 75th percentiles, of the count ratio for ID1 and ID2, we found that the excess emission in ID1 compared to ID2 is detected at a significance level of $4.5\sigma$ (ranging between $4\sigma$ and $5\sigma$). This calculation assumes a similar PSF for both sources, which is reasonable given their proximity.

Figure~\ref{fig:imagesID2datapsf} displays the 0.5-2 keV images for both the observational data (left) and the simulated PSF (right). This comparison highlights the excess counts beyond 2$\arcsec$ around QSO ID1, as also shown in the left panel of Figure~\ref{fig:rp2}.

\section{Proof of equation (\ref{eq:tchia})}\label{app:eqchi}
Equation (\ref{eq:tchia}) can be derived for instance from 3.241-4 in \cite{Gradshteyn07}.  However, the original source, containing the mathematical proof, is of uneasy retrieval. Furthermore, it is unclear from \cite{Gradshteyn07} whether the formula has been proven for non-integer values of $a$. We therefore provide here a simple, explicit proof of equation (\ref{eq:tchia}), valid for all $a>0.5$:
\begin{equation} \label{eq:chi}
\tilde{\chi}(a) = \int_0^{+\infty} (1+x^2)^{-a} dx = \frac{1}{2} \int_0^1 y^{a-\frac{3}{2}} (1-y)^{-\frac{1}{2}} dy = \frac{1}{2} B\left (a - \frac{1}{2}, \frac{1}{2} \right )
= \frac{\sqrt{\pi}}{2} \; \dfrac{ \Gamma \left( a - \dfrac{1}{2} \right)}{\Gamma(a)} ,
\end{equation}
where we used the substitution $y=1/(1+x^2)$, then the definition of Euler's beta function $B$ and finally the well known properties $B(z_1,z_2) = \Gamma(z_1)\Gamma(z_2)/\Gamma(z_1+z_2)$ and $\Gamma(\frac{1}{2}) = \sqrt{\pi}$. Note that for $a\leq0.5$ the integral diverges and $\tilde{\chi}(a)$ is undefined, or formally infinite.

\section{Brunt-V\"ais\"al\"a time-scale for an isothermal $\beta$ model}\label{app:tBV}

Following \cite{Wibking25}, we can define the Brunt-V\"ais\"al\"a time-scale as:
\begin{equation}
    t_{\text{BV}} = \sqrt{\gamma\frac{h_S}{g}} \; ,
\end{equation}
where $g$ in the gravitational field and $h_S$ is the local entropy scale-length defined as $h_S = (dS/dr)^{-1}$, with $S= \ln (T n^{1-\gamma})$. For an isothermal $\beta$-model and assuming $\gamma = 5/3$ (for a monoatomic gas), we find:
\begin{equation}
    h_S (r) = \frac{r^2+r_c^2}{2 \beta r}
\end{equation}
and therefore:
\begin{equation}
t_{\text{BV}}(r) = \sqrt{\frac{5(r^2+r_c^2)}{6 \beta r g(r)}} \; .
\end{equation}

\end{document}